# Extended Kohler's Rule of Magnetoresistance

Jing Xu[1,2,+], Fei Han[3,+], Ting-Ting Wang[4,+], Laxman R. Thoutam[1,5], Samuel E. Pate[1,6], Mingda Li[3,*], Xufeng Zhang[2], Yong-Lei Wang[4,*], Roxanna Fotovat[1], Ulrich Welp[1], Xiuquan Zhou[1], Wai-Kwong Kwok[1], Duck Young Chung[1], Mercouri G. Kanatzidis[1,7], and Zhi-Li Xiao[1,6,*]

[1]*Materials Science Division, Argonne National Laboratory, Argonne, Illinois 60439, USA*
[2]*Center for Nanoscale Materials, Argonne National Laboratory, Argonne, Illinois 60439, USA*
[3]*Department of Nuclear Science and Engineering, Massachusetts Institute of Technology, Cambridge, Massachusetts 02139, USA*
[4]*Research Institute of Superconductor Electronics, School of Electronic Science and Engineering, Nanjing University, Nanjing 210093, China*
[5]*Department of Electronics and Communications Engineering, SR University, Warangal Urban, Telangana 506371, India*
[6]*Department of Physics, Northern Illinois University, DeKalb, Illinois 60115, USA*
[7]*Department of Chemistry, Northwestern University, Evanston, Illinois 60208, USA*

A notable phenomenon in topological semimetals is the violation of Kohler's rule, which dictates that the magnetoresistance $MR$ obeys a scaling behavior of $MR = f(H/\rho_0)$, where $MR = [\rho(H)-\rho_0]/\rho_0$ and $H$ is the magnetic field, with $\rho(H)$ and $\rho_0$ being the resistivity at $H$ and zero field, respectively. Here we report a violation originating from thermally-induced change in the carrier density. We find that the magnetoresistance of the Weyl semimetal, TaP, follows an extended Kohler's rule $MR = f[H/(n_T\rho_0)]$, with $n_T$ describing the temperature dependence of the carrier density. We show that $n_T$ is associated with the Fermi level and the dispersion relation of the semimetal, providing a new way to reveal information on the electronic bandstructure. We offer a fundamental understanding of the violation and validity of Kohler's rule in terms of different temperature-responses of $n_T$. We apply our extended Kohler's rule to $BaFe_2(As_{1-x}P_x)_2$ to settle a long-standing debate on the scaling behavior of the normal-state magnetoresistance of a superconductor, namely, $MR \sim tan^2\theta_H$, where $\theta_H$ is the Hall angle. We further validate the extended Kohler's rule and demonstrate its generality in a semiconductor, InSb, where the temperature-dependent carrier density can be reliably determined both theoretically and experimentally.

# I. Introduction

The magnetic-field-induced resistance change is conventionally termed as magnetoresistance (MR)[1]. In 1938, Max Kohler[2] developed a rule to account for the magnetoresistances in metals. Kohler's rule states that the magnetoresistance $MR$ should be a function of the ratio $H/\rho_0$ of the magnetic field $H$ to the zero-field resistivity $\rho_0$. That is, the field dependence of the magnetoresistances should exhibit a scaling behavior of $MR = f(H/\rho_0)$, where $MR = [\rho(H)-\rho_0]/\rho_0$ with $\rho(H)$ and $\rho_0$ being the resistivity at $H$ and zero field at a fixed temperature, respectively. Kohler's rule of the magnetoresistances has been observed in materials[3-15] beyond metals and recently has been extensively used to understand novel magnetoresistance behavior such as the 'turn-on' temperature behavior of the magnetoresistance in topological materials[5-15]. Violations of Kohler's rule have been often reported[16-40] and used as evidence for phase transitions[16,17] or for emergent new physics[18-22]. Here we explore Kohler's rule of magnetoresistances in Weyl semimetals, where both its validity[5,6,8] and violation[23-25,27,28] have been reported, with the aim to uncover the role played by the inevitable thermally induced change in the density of carriers on the scaling behavior. We find that the magnetoresistance of the Weyl semimetal, tantalum phosphide (TaP), follows an extended Kohler's rule $MR = f[H/(n_T\rho_0)]$, where $n_T$ describes the relative change induced by thermal excitation in the carrier density, with $n_T = 1$ denoting the original Kohler's rule. We outline an innovative approach to obtain $n_T$ without knowing the values of the carrier density, providing a new way to probe the temperature dependence of carrier density. We show that $n_T$ is associated with the Fermi level and the dispersion relation, thereby revealing information on the electronic bandstructure. We offer a fundamental description of the violation and validity of Kohler's rule in terms of different temperature responses of $n_T$. In particular, Kohler's rule is expected to be violated in materials with low carrier density where a noticeable density change due to thermal excitation occurs, while Kohler's rule should hold in materials where the carrier density is high enough such that its thermally-induced change is experimentally indiscernible. Furthermore, by investigating the scaling behavior of the normal-state magnetoresistance in superconducting BaFe$_2$(As$_{1-x}$P$_x$)$_2$ crystals, we demonstrate that our extended Kohler's rule can account for the scaling $MR \sim tan^2\theta_H$ where $\theta_H$ is the Hall angle, which was first observed in cuprate superconductors $YBa_2Cu_3O_{7-\delta}$ and $La_2Sr_xCuO_4$ [18] and has led to a long-standing debate in describing the normal-state magnetoresistance of a superconductor[19-21,33-41]. We also demonstrate the generality of our extended Kohler's rule in an undoped narrow-gap semiconductor, indium antimonide (InSb). At high temperatures, InSb is a compensated two-band system with high-mobility electrons and low-mobility holes, enabling us to determine the carrier density via Hall measurements and use its temperature dependence to test the validity of the $n_T$ term in our extended Kohler's rule.

## II. MATERIALS AND METHODS

Data reported here were obtained from two TaP crystals (sample TP1 and sample TP2), two $BaFe_2(As_{1-x}P_x)_2$ crystals with $x$ = 0.25 (sample PL) and $x$ = 0.5 (sample PH), and one undoped InSb crystal (sample IS).

**Preparation of samples.** (1) TaP: Centimeter-sized single crystals of TaP were grown using the vapor transport method in two steps. In the first step, 3 grams of Ta (Beantown Chemical, 99.95%) and P (Beantown Chemical, 99.999%) powders were weighed, mixed, and ground in a glovebox. The mixed powders were sealed in an evacuated quartz tube which was subsequently heated to 700°C and sintered for 20 hours for a pre-reaction. In the second step, the obtained TaP powder along with 0.4 grams of $I_2$ (Sigma Aldrich, ≥99.8%) were sealed in a new evacuated quartz tube, and was subsequently placed in a two-zone furnace with zone temperatures of 900°C and 950°C, respectively. The crystal growth time was 14 days. Due to the very high electrical conductivity of TaP, it is difficult to carry out high-precision electrical measurements on the as-grown crystals. To increase the electrical resistance of the samples, we polished the crystals down to a thin slab along the c-axis (thickness of tens of micrometers). Electrical leads were gold wires glued to the crystals using silver epoxy H20E. (2) $BaFe_2(As_{1-x}P_x)_2$: crystals of $BaFe_2(As_{1-x}P_x)_2$ with doping levels of $x$ = 0.25 and $x$ = 0.50 were grown using the self-flux method[42]. High-purity flakes of Ba (99.99%, Aldrich) and powders of FeAs and FeP (homemade from Fe, As, and P, 99.99%,

Aldrich) were thoroughly mixed and placed in an $Al_2O_3$ crucible, which was then sealed in an evacuated quartz tube under vacuum and placed in a Lindberg box furnace. Crystals of plate shapes with lateral dimensions up to 2 mm and thicknesses up to 200 µm were obtained by heating up to 1180 °C and then cooling down to 900 °C at a rate 2 °C/min. Electrical leads of gold wires with a diameter of 50 µm were attached to the crystals using silver epoxy H20E. (3) InSb: a crystal of 5 mm × 5 mm × 0.5 mm was purchased from the MTI corporation. It was cut into pieces with desired lateral dimensions. Gold pads of ~100 nm thick were deposited on locations pre-defined using photolithography. Electrical leads were fabricated by attaching 50 µm diameter gold wires to the pads with silver epoxy H20E.

**Resistance measurements.** We conducted resistance measurements to obtain both $R_{xx}(H)$ and $R_{xy}(H)$ curves at various fixed temperatures, enabling the calculation of the resistivities $\rho_{xx}(H) = R_{xx}wd/l$ and $\rho_{xy}(H) = R_{xy}d$, where $w$, $d$, and $l$ are the width, thickness of the sample and the separation between the voltage contacts, respectively. The magnetic field is applied along the *c*-axis of the crystals. The magnetoresistance is defined as $MR = [\rho_{xx}(H) - \rho_0)]/\rho_0$, where $\rho_{xx}(H)$ and $\rho_0$ are the resistivities at a fixed temperature with and without the presence of a magnetic field, respectively. In some cases, we obtained $\rho_{xx}(T)$ curves from the measured $\rho_{xx}(H)$ curves at fixed temperatures to avoid nonequilibrium temperature effects. Data for TaP and InSb were obtained using the conventional four-probe DC electrical transport measurement technique while those for $BaFe_2(As_{1-x}P_x)_2$ were obtained using a low-frequency lock-in method.

### III. RESULTS AND DISCUSSION

**III.1. Extended Kohler's rule of magnetoresistance in Weyl semimetal TaP**

TaP is a transition-metal monophosphide considered as the first realization of a Weyl semimetal[43-46]. Figure 1a presents the typical magneto-transport behavior of the longitudinal resistivity, $\rho_{xx}(H)$, of a TaP crystal (sample TP1). Since the temperature dependence of $\rho_{xx}(T)$ has often exhibited interesting magnetoresistance phenomena such as the 'turn-on' temperature behavior[5,7,8] and topological insulating states[28-30], we also present $\rho_{xx}(T)$ curves in Fig.1b.

Both $\rho_{xx}(H)$ and $\rho_{xx}(T)$ in Fig.1 are consistent with those reported recently in other materials with extremely large magnetoresistance[5-16,22-32]. The magnetoresistance $MR$ can be as high as $10^5$ % at 2 K and 9 T (Fig.2a), and follows the typical power-law $MR \sim H^m$ behavior with $m \approx 1.6$ (dashed line in Fig.2c)[27]. In the absence of a magnetic field, the resistivity decreases monotonically with temperature, as expected for a semimetal. When an external magnetic field is applied, the resistivity increases and a remarkable 'turn-on' behavior appears in the $\rho_{xx}(T)$ curves: the temperature behavior of the resistivity changes from metal-like at high temperatures to insulating-like at low temperatures at magnetic fields $H$ = 0.2 T, 0.5 T and 1.0 T and become insulating-like over the entire temperature range at higher magnetic fields.

The key results of this work on TaP are displayed in Fig.2, which presents the violation of Kohler's rule and highlights our extended Kohler's rule. For clarity, we present in Fig.2a partial $MR(H)$ curves derived from the $\rho_{xx}(H)$ data in Fig.1a while all $\rho_{xx}(T)$ curves in Fig.1b were used to obtain the $MR(T)$ curves in Fig.2d. The respective scaling of $MR(H)$ and $MR(T)$ is presented in Fig.2b and 2e. Clearly, Kohler's rule $MR = f(H/\rho_0)$ is not followed in TaP since the $MR$ curves do not collapse onto a single curve when plotted against $H/\rho_0$. However, all curves in Fig.2b are nearly in parallel with each other, suggesting that a single temperature-dependent multiplier to $MR$ ($y$-axis) or to $H/\rho_0$ ($x$-axis) could cause them to overlap or collapse onto the same curve. Here we tackle the latter case and uncover the underlying physics.

For the convenience of the forthcoming discussions we designate $1/n_T$ as the temperature dependent multiplier to $H/\rho_0$ ($x$-axis) in the $MR \sim H/\rho_0$ curves in Fig.2b. In practice, we scale all the $MR$ curves to the $T$ = 300 K curve [i.e., $n_T$ = 1 for $MR(300K) \sim (1/n_T)(H/\rho_0)$] by varying $n_T$ for each curve. As shown in Fig.2c, all curves in Fig.2b indeed collapse onto the $T$ = 300 K curve when the data are scaled as $MR \sim H/(n_T\rho_0)$. The $n_T$ values for the $MR \sim H/\rho_0$ curves at various temperatures are presented in Fig.3. It decreases monotonically from $n_T$ = 1 at $T$ = 300 K to $n_T$ = 0.45 at $T$ = 2 K. As presented in Fig.2f, we also accounted for the violation of Kohler's rule in Fig.2e for the $MR(T)$ curves by using the same $n_T$ values as those used in Fig.2c. We found nearly identical behavior for the second TaP crystal (sample TP2), as demonstrated by the plots of $MR \sim H/(n_T\rho_0)$ in Fig.S1, with $n_T$ presented in Fig.3. That is, the magnetoresistance of TaP follows an extended Kohler's rule with a thermal factor $n_T$,

$$MR = f[H/(n_T\rho_0)] \qquad (1)$$

In the analysis above we purposely used $n_T$ in the denominator to couple with $\rho_0$ in Eq.1, aiming to reveal the possible role of the carrier density in the violation of Kohler's rule. In a theoretical consideration on the violation of Kohler's rule in the normal-state resistivity of cuprate superconductors, Luo et.al.[41] introduced a MR scaling form (it was called a modified Kohler's rule), $MR = f(H\tau)$, where $\tau$ is the relaxation time of the carriers, since $H$ and $\tau$ are coupled together as $H\tau$ for MRs in the derived equations. Using $\rho_0 = m^*/(ne^2\tau)$, $H\tau$ can be expressed as $H\tau = (m^*/e^2)H/(n\rho_0)$, indicating that a scaling behavior of $MR = f[H/(n\rho_0)]$ is expected if $m^*$ is temperature independent. However, there is no reliable way to obtain the values of $n$ in a semimetal from transport measurements, as demonstrated in Fig.S2 for $n_H/n$ in a compensated system with $n_e = n_h = n$, where $n_H$ is the density obtained from Hall effect measurements, and in Fig.S3 for electron and hole density derived from a two-band model analysis in TaP. Furthermore, $m^*$ can change with temperature[47]. More importantly, materials such as multi-band semimetals where violation of Kohler's rule are reported, go beyond the (isotropic) single-band consideration in $MR = f(H\tau)$. We reveal in the discussion below that $H$ and the mobility $\mu_i = e\tau/m_i^*$ ($i \geq 1$) are interconnected as $H\mu_i$ in the $MR$ expression. Also, it is the temperature-induced change in the carrier density is responsible for the violation of Kohler's rule. Hence, our extended Kohler's rule expression in Eq.1 provides a unique way to reveal the temperature dependence of the carrier density.

In the normal state of a cuprate superconductor the temperature-induced change in the carrier density can be attributed to the existence of a pseudogap or Mott − Wannier excitons of weakly bound electrons and holes[41]. Differing from the linear temperature dependence of the carrier density used to explain the violation of Kohler's rule in cuprate superconductors[41], the $n_T$ obtained in our samples can be approximately described by $n_T \sim T^2$ (Fig.S4), which can be attributed to the thermally-induced change in the carrier density, as elaborated below.

As revealed by bandstructure calculations and angle-resolved photoemission spectroscopy (ARPES) experiments[43-46], TaP has a total of 12 pairs of Weyl nodes. Four of them, denoted as W1, are enclosed by electronlike Fermi surface with energies below the Fermi level. The other eight pairs, denoted as W2, are enclosed by holelike Fermi surface and with energies above the Fermi level. The bandstructure is illustrated in Fig.3, with the relative locations of W1 and W2 with respect to the Fermi level, $E_F$. Considering the temperature dependent Fermi-Dirac distribution, the total density of the conduction electrons at a given $T$ can be straightforwardly obtained using:

$$n = \sum_i \int_{E_i}^{\infty} g_i d\varepsilon / [1 + e^{(\varepsilon - E_F)/k_B T}] \qquad (2)$$

where $i$ = 1 and 2 correspond to Weyl nodes W1 and W2 with respective energies of $E_i$ = $E_1$ and $E_2$. $g_i = g_1(\varepsilon)$ and $g_2(\varepsilon)$ are the density of states (DoS) for the conduction bands of W1 and W2, respectively. Since equal number of holes are created, Eq.2 can also be used to calculate the thermally-induced change in the hole density. In TaP, the density of electron and holes are close to each other (Fig.S3c). We focus only on electron density in the following discussions.

The energies $E_1$ and $E_2$ of the Weyl nodes W1 and W2 in TaP have been determined by bandstructure calculations[44] and by ARPES measurements[45]. The theoretical DoS roughly follows $g \sim \varepsilon$ at $\varepsilon$ up to ~0.3 eV and becomes nearly constant at higher energies[44,47]. As an estimate we use $g_i = g_{i0}(\varepsilon - E_i)$ for $\varepsilon \leq 0.3$ eV and $g_i = g_{i0}(0.3 - E_i)$ for higher energies. In Fig.S5a we present the calculated $n_1$ and $n_2$ and their sum $n$ using the theoretical values of (-53.1 meV, 19.6 meV) for the energies ($E_1$, $E_2$) (relative to the Fermi level) of the Weyl nodes W1 and W2 and the electrons and hole densities of $n_{e0}$ = 4.898×10$^{24}$ m$^{-3}$ and $n_{h0}$ = 5.317×10$^{24}$ m$^{-3}$ at $T$ = 0 K [44]. As indicated by the dashed curve in Fig.3, the total electron density, $n$, normalized to the value at $T$ = 300 K, describes the experimental $n_T$ very well. The small deviation can be understood with sample-dependent ($E_1$, $E_2$), as demonstrated by the dotted curve obtained using the experimental values[45] of (-40 meV, 24 meV) and the solid curve calculated with (-46.2 meV, 19.6 meV), i.e., a slight increase in the theoretical $E_1$ value towards the experimental one while keeping the value of $E_2$ unchanged.

The above discussion indicates that the relative position of the Fermi level to the bottom of the conduction band and the top of the valence band, i.e., the density of electrons and holes at $T = 0$ K affects the temperature dependence of $n_T$. As presented in Fig.S5b, the temperature-induced change in $n_T$ mostly comes from the electron band. In Fig.S6a we present $n_T$ versus $T$ curves calculated using different values of $E_F$ in the electron band. It is clear that the sensitivity of $n_T$ to $T$ depends strongly on the Fermi level. In our TaP samples with $E_F \approx 50$ meV to the bottom of the conduction band, $n_T$ nearly doubles when the temperature is increased from $T = 2$ K to 300 K. However, it becomes challenging to experimentally resolve the change in $n_T$ in the same temperature range when the Fermi level is increased to 200 meV. In this case, $n_T \approx 1$ and Kohler's rule should hold within experimental errors. Thus, the Fermi energy, i.e., electron density at $T = 0$ K, plays a key role in Kohler's rule. It directly explains why Kohler's rule is violated in type-I Weyl semimetals while it is upheld in their type-II counterparts since the latter typically have much higher electron densities[5,6,8]. The validity of Kohler's rule in conventional metals is also not a surprise. They have much higher Fermi energies of a few eVs and electron densities of $10^{28 \sim 29}$ m$^{-3}$ [48], which is 3~4 orders of magnitude higher than that of TaP, making thermally induced changes in the carrier density irrelevant. In Fig.S6b we show that the functional form of the DoS $g(\varepsilon)$ can further contribute to the $n_T$ versus $T$ relationship. When the exponent $\alpha$ in $g(\varepsilon) \sim \varepsilon^\alpha$ changes from $\alpha = 1$ for TaP to $\alpha = 1/2$ for typical metals, $n_T$ becomes less sensitive to the change of $T$ at the same Fermi level. This explains why violation of Kohler's rule is observed more often in topological semimetals than in their trivial counterparts.

**III.2. Extended Kohler's rule versus other alternative scaling forms of the magnetoresistance**

In addition to topological materials, violations of Kohler's rule were often reported in cuprates and iron-based superconductors as well as other topologically trivial materials and various alternative *MR* scaling forms have been introduced[18-20,33-41]. Among them, the most common one is

$$MR = \gamma_H tan^2\theta_H \qquad (3)$$

where $\theta_H = arctan(\rho_{xy}/\rho_{xx})$ is the Hall angle. This MR scaling behavior was first reported in cuprate superconductors, with $\gamma_H$ = 1.7 and $\gamma_H$ = 1.5-1.7 for underdoped and optimally doped YBa$_2$Cu$_3$O$_{7-\delta}$, respectively, and $\gamma_H$ = 13.6 for La$_2$Sr$_x$CuO$_4$ [18]. It has numerous explanations[18-21,37-41] including the spin-charge separation scenario of the Luttinger liquid[21], current vertex corrections and spin density wave[20]. Below we show that the scaling Eq.3 is a natural outcome of our extended Kohler's rule Eq.1 in a compensated two-band material when the carrier mobility is very low, with $\gamma_H$ being an indicator of the ratio of the hole and electron mobility. We also obtain similar scaling forms to Eq.1 and Eq.3 for a low-mobility single-band system with an anisotropic Fermi surface as well as non-compensated two-band and multi-band systems.

For a compensated two-band system where $n_e = n_h = n$, the second term in the denominator of Eq.S1 is zero, leading to $MR = \mu_e\mu_h H^2$. Hence MR can be re-written as $MR = \alpha_\mu/e^2[H/(n\rho_0)]^2$ where $\alpha_\mu = (\mu_h/\mu_e)/(1+\mu_h/\mu_e)^2$, $\rho_0 = [ne(\mu_h+\mu_e)]^{-1}$ and $n_e$, $n_h$, $\mu_e$ and $\mu_h$ are the densities and mobilities of electron and holes, respectively. Thus, our extended Kohler's rule presented in Eq.1 is valid if $\alpha_\mu$ is temperature independent. Likewise, the second term in the numerator of Eq.S2 also becomes zero, yielding a linear Hall resistivity $\rho_{xy} = H(\mu_h/\mu_e-1)/[en(\mu_h/\mu_e+1)]$. Then, $MR = \mu_e\mu_h H^2$ can be alternately expressed as

$$MR = \gamma_H(\rho_{xy}/\rho_0)^2 \quad (4)$$

with
$$\gamma_H = (\mu_h/\mu_e)/(1-\mu_h/\mu_e)^2 \quad (5)$$

At very low mobilities, $MR \rightarrow 0$ and $\rho_{xx} \approx \rho_0$, resulting in Eq.3 and Eq.4 to be equivalent. That is, in a compensated two-band system with very low mobilities, our extended Kohler's rule in Eq.1 will lead to the scaling behavior of Eq.3.

The same conclusion can be reached for a nearly compensated two-band system, if the mobilities and/or the magnetic fields are low, such that the second term in the denominator of both Eq.S1 and Eq.S2 as well as the second term in the numerator of Eq.S2 become negligible. As revealed in TaP below, $\alpha_\mu$ is indeed temperature insensitive (Fig.5b). Thus, our extended Kohler's rule Eq.1 can be expressed as Eq.3 for a compensated system as well as for a nearly compensated system with very low mobilities and/or at very low fields if the mobilities are high.

Underdoped $YBa_2Cu_3O_{7-\delta}$ [49,50] are indeed two-band materials. On the other hand, the optimal-doped $YBa_2Cu_3O_{7-\delta}$ (with $T_c$ = 90 K) which also shows the scaling behavior akin to Eq.3 is believed to be single-band system with an anisotropic Fermi surface[41]. As detailed in the supplement, a single band material would have no magnetoresistance if all carriers moving in the same direction have the same mobility. It exhibits magnetoresistance probably due to (1) the mobility distribution of carriers from different energy levels near the Fermi surface[1] and (2) the existence of impurities. In the first case, we can obtain $MR \approx \alpha_\mu^*[H/(n\rho_0)]^2$ (Eq.S10) with $\alpha_\mu^* = 2\kappa/e^2[1 + (\mu_L/\mu_H)^2]/[1 + \mu_L/\mu_H]^2$ (Eq.S11) at low carrier mobilities such that $\mu_x^2 H^2 \ll 1$, where $\mu_H$ and $\mu_L$ represent the highest and lowest mobility of $\mu_x$ along the x-direction and $\kappa$ is the ratio $(\mu_y/\mu_x)$ of the carrier mobilities $\mu_x$ and $\mu_y$ along the $x$ and $y$ directions. In the meantime, we can also obtain $MR \approx \gamma_H^* tan^2\theta_H$ with $\gamma_H^* = (9/8\kappa)[1 + (\mu_L/\mu_H)^2][1 - (\mu_L/\mu_H)^2]^2/[1 - (\mu_L/\mu_H)^3]^2$ (Eq.S12). In the latter case, an optimal-doped $YBa_2Cu_3O_{7-\delta}$ crystal can be considered as a two-band or multiband system, with a dominating intrinsic anisotropic band together with one or more impurity bands. At low carrier mobilities, we have $MR = [H/(n_T\rho_0)]^2$ (Eq.S15) with $n_T = e[\sum_i(n_i\mu_i)]^{3/2}/[\sum_i(n_i\mu_i^3)]^{1/2}$ (Eq.S16), and $MR \approx \gamma_H^{**}(\rho_{xy}/\rho_0)^2$ (Eq.S17) with $\gamma_H^{**} = \sum_i(n_i\mu_i^3)\sum_i(n_i\mu_i)/[\sum_i(n_i\mu_i^2)]^2$, $\rho_0 = 1/\sum_i(en_i\mu_i)$, $n_i$ and $\mu_i$ being the carrier density and mobility of the i$^{th}$ band. These results not only explain the observed scaling behavior of Eq.3 in optimal-doped $YBa_2Cu_3O_{7-\delta}$ but also indicate that our extended Kohler's rule Eq.1 should be valid whenever the scaling of Eq.3 is observed.

TaP is a nearly compensated two-band system with high mobilities, as manifested by the large MRs (Fig.2a) and non-linear $\rho_{xy}$ curves (Fig.S7). Thus, the scaling behavior in Eq.3 is expected to fail when the MRs become significantly large, as confirmed by the plot in Fig.4a, which shows that Eq.3 is roughly valid for *MR* < 10% and yields a value of $\gamma_H$ = 9 at *T* = 300 K. On the other hand, Eq.4 is an approximate expression of Eq.1 at low magnetic fields and should be valid over a wider field range by avoiding the influence of $\rho_{xx}(H)$ in Eq.3. As plotted in Fig.4b, $MR \sim (\rho_{xy}/\rho_0)^2$ indeed allows us to more reliably derive the $\gamma_H$ values (Fig.5a). The corresponding $\mu_h/\mu_e$ changes from 1.39 at $T$ = 300 K to 1.24 at $T$ = 2 K (Fig.5b), leading to a nearly temperature-independent $\alpha_\mu$ with a very small change (< 0.35%) from $T$ = 300 K to $T$ = 2 K (Fig.5b).

In both Kohler's theory[2] and the derivations by Luo et al.[41], $H\tau$ appears as a product that is inseparable in the expression for $MR = f(H\tau)$ and hence has been proposed as a modified Kohler's rule[33,36,41]. The magnetoresistances in the normal state of La$_{2-x}$Sr$_x$CuO$_4$ and K$_x$Fe$_{2-y}$Se$_2$ single crystals were indeed found to follow the scaling behavior of $MR = f(H\tau)$ if $\tau \sim T^{-1}$ [33] and $\tau \sim T^{-2}$ [36] are respectively assumed. However, $\tau$ is not a parameter that can be conveniently obtained from resistivity measurements. As discussed in section III.1, $MR = f(H\tau)$ does not consider the possible role of the carrier's effective mass and $\tau$ should be replaced with $\mu$, which can have more than one value, as demonstrated in Eq.S10 and Eq.S11 as well as Eq.S15 and Eq.S16. Thus, its applications are limited, particularly for single band systems with temperature independent effective mass. It will usually fail in two-band and multi-band systems. For example, by assuming single relaxation time $\tau$ for all carriers we can re-write Eq.S15 as $MR = f(\tau_T H \tau)$ with $\tau_T = \sum_i (n_i/m_i^3)]^{1/2} / \sum_i (n_i/m_i)]^{1/2}$. Clearly, $\tau_T$ will not be a constant when $n_i$ and $m_i$ are temperature dependent. That is, Eq.1 extends the Kohler's rule to two-band and multi-band systems, as experimentally confirmed by the MR scaling behavior in TaP. These derivations further indicate that $n_T$ represents the temperature dependence of the carrier density in an anisotropic single-band system ($n_T \sim n$, see Eq.S10) as well as in two-band and multi-band systems ($n_T \sim f_n(T)$, see Eq.S16) if the densities/mobilities from different bands have the same or similar temperature dependence, i.e., $n_i \approx n_i^0 f_n(T)$ and $\mu_i \approx \mu_i^0 f_\mu(T)$. In other cases, the thermal factor $n_T$ in Eq.1 contains contributions from the temperature dependences of the carrier densities and mobilities of all bands.

### III.3. Generality of the extended Kohler's rule

III.3.1. Extended Kohler's rule of normal-state magnetoresistance in superconductor BaFe$_2$(As$_{1-x}$P$_x$)$_2$

The discussions in the preceding section indicate that Eq. 1, our extended Kohler's rule applied to a semimetal also provides a sensible explanation for the alternative scaling rule presented in Eq.3, which has been used routinely to account for the normal-state magnetoresistances in several classes of superconductors[19-21,33-41]. Here, we experimentally confirm the applicability of Eq.1 on two superconducting BaFe$_2$(As$_{1-x}$P$_x$)$_2$ crystals with $x$ = 0.25 and 0.5, respectively. Their corresponding superconducting transition temperatures are 31 K and 22 K, as revealed by the temperature dependence of the zero-field resistivity presented in Fig.S8.

In the over-doped crystal with $x$ = 0.5, we found that the magnetoresistance obeys Kohler's rule (Fig.S9c) while the plot of $MR$ versus $tan^2\theta_H$ does not collapse the data into a single curve (Fig.S9b). However, as indicated by the dashed magenta line, the $MR$ versus $tan^2\theta_H$ curves obtained at different temperature are indeed linear. The parallel shift in the log-log plot indicates that the prefactor $\gamma_H$ in Eq.3 is temperature dependent (inset of Fig.S9b), similar to that found in TaP (Fig.5a). In the under-doped crystal with $x$ = 0.25, we did observe both the violation of Kohler's rule (Fig.6b) and the validity of the scaling Eq.3 (Fig.6c). As indicated in Fig.6c and the $\gamma_H$ values in its inset, $MR$ versus $tan^2\theta_H$ curves at $T \geq 50$ K overlap each other while those at $T < 45$ K show a slight parallel shift, with an increase of $\gamma_H$ from ~7 at $T$ = 45 K to ~9 at $T$ = 32 K. In contrast, our extended Kohler's rule, Eq.1, works well over the entire temperature range as shown in Fig.6d. Similar to $\gamma_H$, the derived $n_T$ (inset of Fig.6d) also shows a significant change in its temperature dependence at $T \approx 50$ K. At $T \geq 50$ K, the temperature dependence of $n_T$ is roughly linear. Interestingly, it can also be described by $n_T = n_0 + \alpha T e^{-\Delta/k_B T}$ (dashed line in the inset of Fig.6d with $n_0$ = 0.7, $\alpha$ = 5.4×10$^{-3}$, and $\Delta$ = 5.18 meV), analogous to the temperature dependence of the carrier density arising from carriers thermally excited over a pseudogap $\Delta$ in cuprates[51,52]. At $T \leq 45$ K, $n_T$ changes with temperature at a much higher rate. The temperature (~50 K) at which $n_T$ changes its temperature sensitivity is coincident with that of a transition into an antiferromagnetic orthorhombic phase[35] (inset of Fig.S8 and discussion in its caption). While further investigations are needed to account for the temperature behavior of $n_T$ at temperatures above and below ~50 K, the results in Fig.6 demonstrate that our extended Kohler's rule will work when the scaling following Eq.3 is obeyed. This provides experimental support for the conclusion in section III.2 that Eq.1 can be expressed as Eq.3 when the mobilities of the carriers are low, which can be inferred from the negligible MRs in BaFe$_2$(As$_{1-x}$P$_x$)$_2$ (Fig.6, up to 2% at $H$ = 9 T). Furthermore, the above discussions show that $n_T$ can be an indicator of a temperature induced phase transition, if it exhibits a sudden change in the temperature sensitivity.

III.3.2. Extended Kohler's rule of magnetoresistance in semiconductor InSb

Following the discussions in section III.1 and III.2, it can be challenging to account for the temperature behavior of $n_T$ derived from our extended Kohler's rule of Eq.1 in semimetals and in the normal state of a superconductor. In the former case, one needs to know detailed information of the semimetal's electronic bandstructure that can be sample dependent[44,45]. In the normal state of a superconductor, $n_T$ can be governed by more than one mechanism besides the electronic bandstructure, such as a pseudogap. In order to unambiguously validate the extended Kohler's rule of Eq.1, we applied the scaling to the magnetoresistance of an undoped semiconductor. Kohler's rule is presumed to be violated due to the expected exponential temperature dependence of the intrinsic carrier density $n_i \sim T^{3/2}e^{-E_g/2k_BT}$ with $E_g$ being the band gap[48], providing an exemplar system to showcase our extended Kohler's rule Eq.1. Its compensated nature also simplifies the analysis using Eq.S1 and EqS2 of the conventional two-band model, as discussed in section III.2. We chose InSb, which is a narrow-gap semiconductor[53] with resistivities conveniently measurable at around room temperature[54]. Its temperature dependent band gap $E_g$ is available in the literature[53], enabling comparisons of the temperature dependence of $n_T$ with that of the calculated $n_i$ and/or of the reported band gap with that derived from $n_T$. Its large MR[54] (up to ~10$^3$% at $T$ > 250 K) implies high carrier mobility, extending the range of the carrier mobility and enabling the validation of Eq.4, from which scaling Eq.3 is deduced at low carrier mobility. More importantly, we found a very large (> 10$^2$) ratio ($\mu_e/\mu_h$) of the electron ($\mu_e$) and hole ($\mu_h$) mobility in InSb, where the carrier density $n_i$ is practically same as that ($n_H$) obtained from Hall measurements (Fig.S2 and caption). This allows a further verification of $n_T$ using the experimentally determined carrier density.

We present typical $\rho_{xx}(H)$ curves of InSb around room temperature in Fig.7a. We focused on data obtained at $T \geq 240$ K to avoid interference of quantum magnetoresistance that can occur at lower temperatures[54] and the contribution to the magnetoresistance by the residual impurity in the nominally undoped crystal (Fig.S10 and caption). As expected for a semiconductor, $\rho_{xx}$ increases with decreasing temperature (also Fig.S10a). Figure 7c shows that Kohler's rule is violated in InSb. In fact, the curves in the Kohler's rule plot are separated from each other even further, compared to those prior to the scaling (Fig.7b). This is in contrast to those shown in Fig.2a&2b and in Fig.6a&b for a semimetal (TaP) and for a superconductor [$BaFe_2(As_{2-x}P_x)_2$] in the normal state, because their zero-field resistivity $\rho_0$ has opposite temperature dependence to that of the semiconducting InSb. On the other hand, our extended Kohler's rule, Eq.1, can collapse all the data into a single curve (Fig.7d). The temperature dependence of the derived $n_T$ can be well described theoretically (Fig.S11b), unveiling band gaps comparable to those determined from other methods in the literature (inset of Fig.S11b). It is nearly indistinguishable to that of the experimental Hall carrier density $n_H$ obtained from the $\rho_{xy}(H)$ curves (Fig.S11a) as well as carrier density $n_i$ (Fig.S12c) obtained by simultaneous fittings of $\rho_{xx}(H)$ and $\rho_{xy}(H)$ curves using the two-band model (Fig.S12a). These results evidently prove the validity of our extended Kohler's rule Eq.1 in semiconducting InSb, further demonstrating its generality. As presented in Fig.7e, a plot of MR versus $tan^2\theta_H$ can also collapse all data into one common curve, which becomes however nonlinear with increasing magnetic field. This indicates that MR is not proportional to $tan^2\theta_H$, i.e., the scaling provided by Eq.3 is not valid. The reason is that Eq.3 is deducible from Eq.4 only when the carrier mobility is very low so that the magnetoresistance MR is negligible, i.e., $\rho_{xx}(H) \approx \rho_0$. On the other hand, Fig.7f shows that the general form Eq.4 does work well in InSb over the entire field range, confirming that in a compensated two-band system, Eq.4, i.e., $MR = \gamma_H(\rho_{xy}/\rho_0)^2$, can be derived from the extended Kohler's rule Eq.1, regardless of the value of the carrier mobility.

## IV. CONCUDING REMARKS

Since it was proposed more than 80 years ago for orbital magnetoresistances in non-magnetic metals, Kohler's rule has been widely used to account for the magnetoresistance behavior in materials beyond simple metals, including cuprate and iron-based superconductors as well as in topological materials. On one hand, it offers a phenomenological understanding of novel magnetoresistance phenomena such as the 'turn-on' temperature behavior of the magnetoresistance in topological materials. On the other hand, its violations have been often reported, and attributed to temperature-induced phase transitions or other unconventional mechanisms. Stimulated by the widely reported violations of Kohler's rule in the newly discovered topological semimetals, we tackled the ubiquitous thermal induced changes in the carrier density. We used a low carrier density, type-I Weyl semimetal, TaP, to establish an extended Kohler's rule (Eq.1), which takes into account the role played by the temperature dependence of the carrier density. We demonstrated how the extended Kohler's rule naturally reduces to the original Kohler's rule in materials such as metals whose carrier density is so high that the temperature induced change in it is experimentally indistinguishable. We applied our extended Kohler's rule to account for other alternative scaling forms of magnetoresistance, particularly the widely debated scaling behavior of $MR \sim tan^2\theta_H$ (Eq.3) discovered in cuprates and often applied to other superconductors. We showed that Eq.3 can be deduced from our extended Kohler's rule Eq.1 when the carrier mobility is very low. We also conducted measurements on the normal-state magnetoresistance in superconductor $BaFe_2(As_{1-x}P_x)_2$ to demonstrate that Eq.1 is valid when the scaling Eq.3 is observed. We further demonstrated the validation and generality of the extended Kohler's rule by investigating the magnetoresistance in a narrow-gap semiconductor, InSb, whose carrier density is expected to change strongly with temperature and can be determined both theoretically and experimentally.

Our extended Kohler's rule Eq.1 offers a fundamental understanding of the violation and validity of Kohler's rule in terms of different temperature-response of the thermal factor $n_T$, with $n_T = 1$ denoting the original Kohler's rule. The results for TaP and InSb evidently show that $n_T$ represents the temperature dependence of the carrier density, providing an alternative way to reveal information on the electronic bandstructure, e.g., Fermi level (in TaP) and band gap (in InSb). On the other hand, our extended Kohler's rule is inconclusive in understanding the temperature dependence of $n_T$ in $BaFe_2(As_{1-x}P_x)_2$ which is a multi-band system and where other mechanisms such as the pseudogap may also contribute to the thermally induced change in the carrier density. In general, we expect $n_T$ to reflect the temperature dependence of the carrier density in (1) single band (or one dominant band) systems and (2) in two-band and multi-band materials whose carrier density and mobility in all bands have the same or similar temperature dependence, as demonstrated by the experimental results in the two-band systems TaP and InSb as well as the derived formulae of the compensated two-band systems and also noncompensated two-band and multi-band materials. The temperature behavior of the thermal factor $n_T$ depends on that of both carrier density and mobility, if a system with two or more bands has different temperature dependences for the carrier density and mobility in each band. In this case, detailed information on the carrier densities and mobilities of all bands are required to calculate $n_T$, making the comparison of theory and experiments more challenging. Further work on more materials is needed to ultimately determine the limitations of our extended Kohler's rule. We note that other mechanisms can also cause violations of Kohler's rule. This work demonstrates that thermal effects on the carrier density and mobility may need to be considered before new mechanisms are proposed.

**ACKNOWLEDGMENTS**

Experimental design, magneto-transport measurements and data analysis were supported by the U.S. Department of Energy, Office of Science, Basic Energy Sciences, Materials Sciences and Engineering. F. H. and M. D. L. acknowledge support DE-SC0020148 for supporting crystal growth. S. E.

P., R. F. and Z.-L. X. received support by the National Science Foundation (Grant No. DMR-1901843). T. T. W. and Y.-L. W. acknowledges support by the Jiangsu Excellent Young Scholar program (BK20200008) and the National Natural Science Foundation of China (61771235).

[+]Authors contributed equally

[*]Correspondence to: mingda@mit.edu; yongleiwang@nju.edu.cn; xiao@anl.gov

**References**

1. J. M. Ziman, Electrons and phonons: The theory of transport phenomena in solids. Cambridge University Press, Cambridge, UK (2001).

2. M. Kohler, Zur magnetischen widerstandsänderung reiner metalle. *Ann. Phys.* **32**, 211-218 (1938).

3. L. Forro et al. Magnetoresistance of the organic superconductor bis-tetramethyltetraselenafulvalenium perchlorate [(TMTSF)$_2$ClO$_4$]: Kohler's rule. *Phys. Rev. B* **29**, 2839-2842 (1984).

4. M. K. Chan et al. In-plane magnetoresistance obeys Kohler's rule in the pseudogap phase of cuprate superconductors. *Phys. Rev. Lett.* **113**, 177005 (2014).

5. Y. L. Wang et al. Origin of the turn-on temperature behavior in WTe$_2$. *Phys. Rev. B* **92,** 180402(R) (2015).

6. A. Narayanan et al. Linear magnetoresistance caused by mobility fluctuations in n-doped Cd$_3$As$_2$. *Phys. Rev. Lett.* **114**, 117201 (2015).

7. F. Han et al. Separation of electron and hole dynamics in the semimetal LaSb. *Phys. Rev. B* **96**, 125112 (2017).

8. Q. L. Pei et al. Origin of the turn-on phenomenon in Td-MoTe$_2$. *Phys. Rev. B* **96**, 075132 (2017).

9. N. H. Jo et al. Extremely large magnetoresistance and Kohler's rule in PdSn$_4$: A complete study of thermodynamic, transport, and band-structure properties. *Phys. Rev. B* **96**, 165145 (2017).

10. J. Du et al. Extremely large magnetoresistance in the topologically trivial semimetal α-WP$_2$. *Phys. Rev. B* **97**, 245101 (2018).

11. O. Pavlosiuk, P. Swatek, D. Kaczorowski, P. Wiśniewski, Magnetoresistance in LuBi and YBi semimetals due to nearly perfect carrier compensation. *Phys. Rev. B* **97**, 235132 (2018).

12. Y. J. Hu et al. Extremely large magnetoresistance and the complete determination of the Fermi surface topology in the semimetal ScSb. *Phys. Rev. B* **98**, 035133 (2018).

13. A. I. U. Saleheen et al. Evidence for topological semimetallicity in a chain-compound $TaSe_3$. *npj Quantum Mater*. **5**, 53 (2020).

14. Q. Chen et al. Large magnetoresistance and nonzero Berry phase in the nodal-line semimetal $MoO_2$. *Phys. Rev. B* **102**, 165133 (2020).

15. R. Chapai et al. Quantum oscillations with angular dependence in $PdTe_2$ single crystals. *J. Phys.: Condens. Matter* **33**, 035601 (2021).

16. Y. Wu et al. Temperature-induced Lifshitz transition in $WTe_2$. *Phys. Rev. Lett.* **115**, 166602 (2015).

17. S. Roßler et al. Emergence of an incipient ordering mode in FeSe. *Phys. Rev. B* **92**, 060505(R) (2015).

18. M. Harris et al. Violation of Kohler's Rule in the normal-state magnetoresistance of $YBa_2Cu_3O_{7-\delta}$ and $La_2Sr_xCuO_4$. *Phys. Rev. Lett.* **75**, 1391-1394 (1995).

19. A. Narduzzo et al. Possible coexistence of local itinerancy and global localization in a quasi-one-dimensional conductor. *Phys. Rev. Lett.* **98**, 146601 (2007).

20. H. Kontani, Anomalous transport phenomena in Fermi liquids with strong magnetic fluctuations. *Rep. Prog. Phys.* **71**, 026501 (2008).

21. P. W. Anderson, When the electron falls apart. *Phys. Today* **50**, 42-47 (1997).

22. X. Li et al. Pressure-induced phase transitions and superconductivity in a black phosphorus single crystal. *Proc. Natl Acad. Sci. USA* **115**, 9935-9940 (2018).

23. C.-L. Zhang et al. Electron scattering in tantalum monoarsenide. *Phys. Rev. B* **95**, 085202 (2017).

24. A. Wang et al. Large magnetoresistance in the type-II Weyl semimetal $WP_2$. *Phys. Rev. B* **96**, 121107(R) (2017).

25. A. Wang et al. Magnetotransport properties of $MoP_2$. *Phys. Rev. B* **96**, 195107 (2017)

26. L.-L. Sun et al. Crystal growth and magneto-transport properties of α-$ZrSb_2$ and α-$HfSb_2$. *Europhys. Lett.* **120**, 37002 (2017).

27. I. A. Leahy et al. Nonsaturating large magnetoresistance in semimetals. *Proc. Natl Acad. Sci. USA* **115**, 10570-10575 (2018).

28. R. Sankar et al. Crystal growth and transport properties of Weyl semimetal TaAs. *J. Phys.: Condens. Matter* **30**, 015803 (2018).

29. B. Qian et al. Extremely large magnetoresistance in the nonmagnetic semimetal YBi. *J. Mater. Chem. C* **6**, 10020 (2018).

30. V. Harimohan et al. Magneto-resistance in pristine and irradiated $TaAs_2$. *AIP Advances* **9**, 045020 (2019).

31. Q. Niu et al. Nonsaturating large magnetoresistance in the high carrier density nonsymmorphic metal CrP. *Phys. Rev. B* **99**, 125126 (2019).

32. A. Laha et al. Magnetotransport properties of the topological nodal-line semimetal CaCdSn. *Phys. Rev. B* **102**, 035164 (2020).

33. T. Kimura et al. In-plane and out-of-plane magnetoresistance in $La_{2-x}Sr_xCuO_4$ single crystals. *Phys. Rev. B* **53**, 8733-8742 (1996).

34. P. Cheng et al. Hall effect and magnetoresistance in single crystals of $NdFeAsO_{1-x}F_x$ (x=0 and 0.18). *Phys. Rev. B* **78**, 134508 (2008).

35. S. Kasahara et al. Evolution from non-Fermi- to Fermi-liquid transport via isovalent doping in $BaFe_2(As_{1-x}P_x)_2$ superconductors. *Phys. Rev. B* **81**, 184519 (2010).

36. X. Ding, Y. Pan, H. Yang, H. H. Wen, Strong and nonmonotonic temperature dependence of Hall coefficient in superconducting $K_xFe_{2-y}Se_2$ single crystals. *Phys. Rev. B* **89**, 224515 (2014).

37. S. Nair et al. Precursor state to unconventional superconductivity in $CeIrIn_5$. *Phys. Rev. Lett.* **100**, 137003 (2008).

38. Y. Sun et al. Multiband effects and possible Dirac fermions in $Fe_{1+y}Te_{0.6}Se_{0.4}$. *Phys. Rev. B* **89**, 144512 (2014).

39. R. Kumar, S. Singh, S. Nair, Scaling of magnetotransport in the $Ba(Fe_{1-x}Co_x)_2As_2$ series. *J. Phys.: Condens. Matter* **31**, 115601 (2019).

40. N. Maksimovic et al. Magnetoresistance scaling and the origin of H-Linear resistivity in BaFe$_2$(As$_{1-x}$P$_x$)$_2$. *Phys. Rev. X* **10**, 041062 (2020).

41. N. Luo, G. H. Miley, Kohler's rule and relaxation rates in high-T$_c$ superconductors. *Physica C* **371**, 259-269 (2002).

42. C. Chaparro et al. Doping dependence of the specific heat of single-crystal BaFe$_2$(As$_{1-x}$P$_x$)$_2$. *Phys. Rev. B* **85**, 184525 (2012).

43. H. Weng et al. Weyl semimetal phase in noncentrosymmetric transition-metal monophosphides. *Phys. Rev. X* **5**, 011029 (2015).

44. C. -C. Lee et al. Fermi surface interconnectivity and topology in Weyl fermion semimetals TaAs, TaP, NbAs, and NbP. *Phys. Rev. B* **92**, 235104 (2015).

45. S. Y. Xu et al. Experimental discovery of a topological Weyl semimetal state in TaP. *Sci. Adv.* **1**, e1501092 (2015).

46. Z. K. Liu et al. Evolution of the Fermi surface of Weyl semimetals in the transition metal pnictide family. *Nat. Mater.* **15**, 27-31 (2016).

47. S.-I. Kimura et al. Optical signature of Weyl electronic structures in tantalum pnictides TaP$_n$ (P$_n$ = P, As). *Phys. Rev. B* **96**, 075119 (2017).

48. C. Kittel, *Introduction to Solid State Physics*. 8th Edition, John Wiley & Sons (2004).

49. S. E. Sebastian et al. Compensated electron and hole pockets in an underdoped high-T$_c$ superconductor. *Phys. Rev. B* **81**, 214524 (2010).

50. N. Doiron-Leyraud et al. Evidence for a small hole pocket in the Fermi surface of underdoped YBa$_2$Cu$_3$O$_y$. *Nat. Commun.* **6**, 6034 (2015).

51. V. V. Kabanov, J. Demsar, B. Podobnik, D. Mihailovic, Quasiparticle relaxation dynamics in superconductors with different gap structures: Theory and experiments on YBa$_2$Cu$_3$O$_{7-\delta}$. *Phys. Rev. B* **59**, 1497-1506 (1999).

52. L. P. Gor'kov, G. B. Teitel'baum, Interplay of externally doped and thermally activated holes in La$_{2-x}$Sr$_x$CuO$_4$ and their impact on the pseudogap crossover. *Phys. Rev. Lett.* **97**, 247003 (2006).

53. C. L. Littler and D. G. Seiler, Temperature dependence of the energy gap of InSb using nonlinear optical techniques. *Appl. Phys. Lett.* **46**, 986-988 (1985).
54. J. S. Hu and T. F. Rosenbaum, Classical and quantum routes to linear magnetoresistance. *Nat. Mater.* **7**, 697-700 (2008).

**Figure captions**

**Fig.1.** Magnetoresistance of TaP. (a) Magnetic field dependence $\rho_{xx}(H)$ measured at temperatures of $T$ = 2 K, and from 5 K to 150 K in intervals of 5 K and from 160 K to 300 K in intervals of 10 K (from red to purple). (b), Temperature dependence $\rho_{xx}(T)$ constructed from $\rho_{xx}(H)$ in (a) at magnetic fields of $H$ = 0 T, 0.1 T, 0.2 T, 0.5 T and from 1 T to 9 T in intervals of 1 T (from purple to red). The data were taken from sample TP1.

**Fig.2.** Extended Kohler's rule of the magnetoresistance in TaP (sample TP1). (a) and (d), Magnetic field and temperature dependences of the *MR* derived from data in Fig.1a and 1b, respectively. For clarity, Fig.2a presents only partial *MR(H)* curves derived from the $\rho_{xx}(H)$ data in Fig.1a. (b) and (e), Kohler's rule plots of the data in (a) and (d), respectively. (c) and (f), Extended Kohler's rule plots of the MR curves in (a) and (d), respectively. The legends for Fig.2a-2c are on the top right while legends for Fig.2d-2f are on the bottom right.

**Fig.3.** Temperature dependence of $n_T$ for samples TP1 (red circles) and TP2 (green circles) derived from the extended Kohler's rule Eq.1. The dashed, solid and dotted curves are calculated electron densities from Eq.2 with energies ($E_1$, $E_2$) (relative to the Fermi level) of the Weyl nodes W1 and W2 of (-53.1 meV, 19.6 meV), (-46.2 meV, 19.6 meV), and (-40 meV, 24 meV), respectively. To show the temperature dependence rather than their absolute values, all the calculated curves are normalized to the values at $T$ = 300 K. Schematics of the electronic bandstructures are presented on the top of the panel.

**Fig.4.** Alternative scalings for the magnetoresistance in TaP (sample TP1) with Eq.3 (a) and Eq.4 (b). The values of $\rho_{xx}$, $\rho_0$ and *MR*s were taken from the same dataset as those used for the extended Kohler's rule in Fig.2a-2c and the corresponding $\rho_{xy}$ taken from Fig.S7a. The solid pink line in (a) and (b) respectively represent $MR = \gamma_H tan^2\theta_H$ and $MR = \gamma_H(\rho_{xy}/\rho_0)^2$ with $\gamma_H$ = 9. The same colored open symbols are used in both panels.

**Fig.5.** Parameters derived from scaling magnetoresistances in Fig.4b using Eq.4. (a) Temperature dependence of $\gamma_H$, obtained at low magnetic fields from the plot of $MR \sim (\rho_{xy}/\rho_0)^2$. (b) The ratio $\mu_h/\mu_e$ of the mobilities and the prefactor $\alpha_\mu$ in $MR = \alpha_\mu/e^2[H/(n\rho_0)]^2$ at various temperatures, where $\mu_h/\mu_e$ is calculated from Eq.5 using the $\gamma_H$ values in (a) and $\alpha_\mu$ is obtained using $\alpha_\mu = (\mu_h/\mu_e)/(1+\mu_h/\mu_e)^2$.

**Fig.6.** Scaling behavior of the magnetoresistance of a BaFe$_2$(As$_{1-x}$P$_x$)$_2$ crystal with $x = 0.25$ (sample PL). (a) $MR(H)$ curves at various temperatures. (b) Scaling according to the Kohler's rule. The red line represents $MR \sim (H/\rho_0)^2$. (c) Scaling according to Eq.3. The dashed straight blue line is $MR = \gamma_H tan^2\theta_H$ with $\gamma_H = 3$, demonstrating the validity of Eq.3. The value of $\gamma_H$ for each temperature is presented in the inset. (d) Scaling according to the extended Kohler's rule Eq.1. The red line represents $MR \sim [H/(n_T\rho_0)]^2$. The inset shows the derived $n_T$, where the dotted purple line describes a possible pseudogap temperature behavior of $n_T = n_0 + \alpha Te^{-\Delta/k_BT}$ with $n_0 = 0.7$, $\alpha = 5.4\times10^{-3}$, and $\Delta = 5.18$ meV (or $\Delta/k_B = 60$ K). The same colored open symbols are used in all panels.

**Fig.7.** Scaling behavior of the magnetoresistance in an InSb crystal (sample IS). (a) $\rho_{xx}(H)$ curves at various temperatures. (b) $MR(H)$ curves at various temperatures. (c) Kohler's rule plot. (d) Scaling according to the extended Kohler's rule Eq.1. (e) and (f) Scaling according to Eq.3 and Eq.4, respectively. The same colored open symbols are used in all panels.

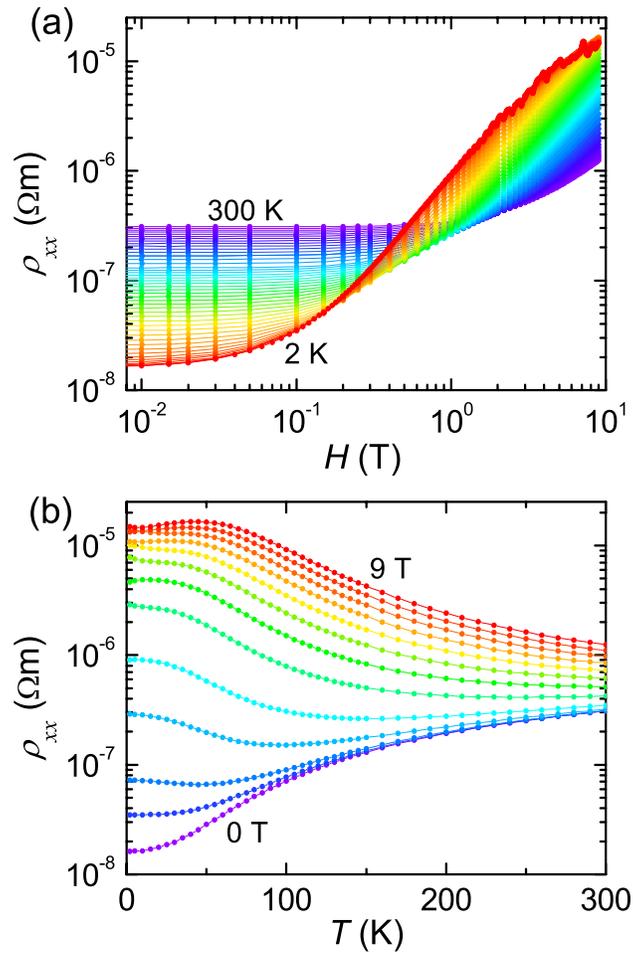

**Figure 2**

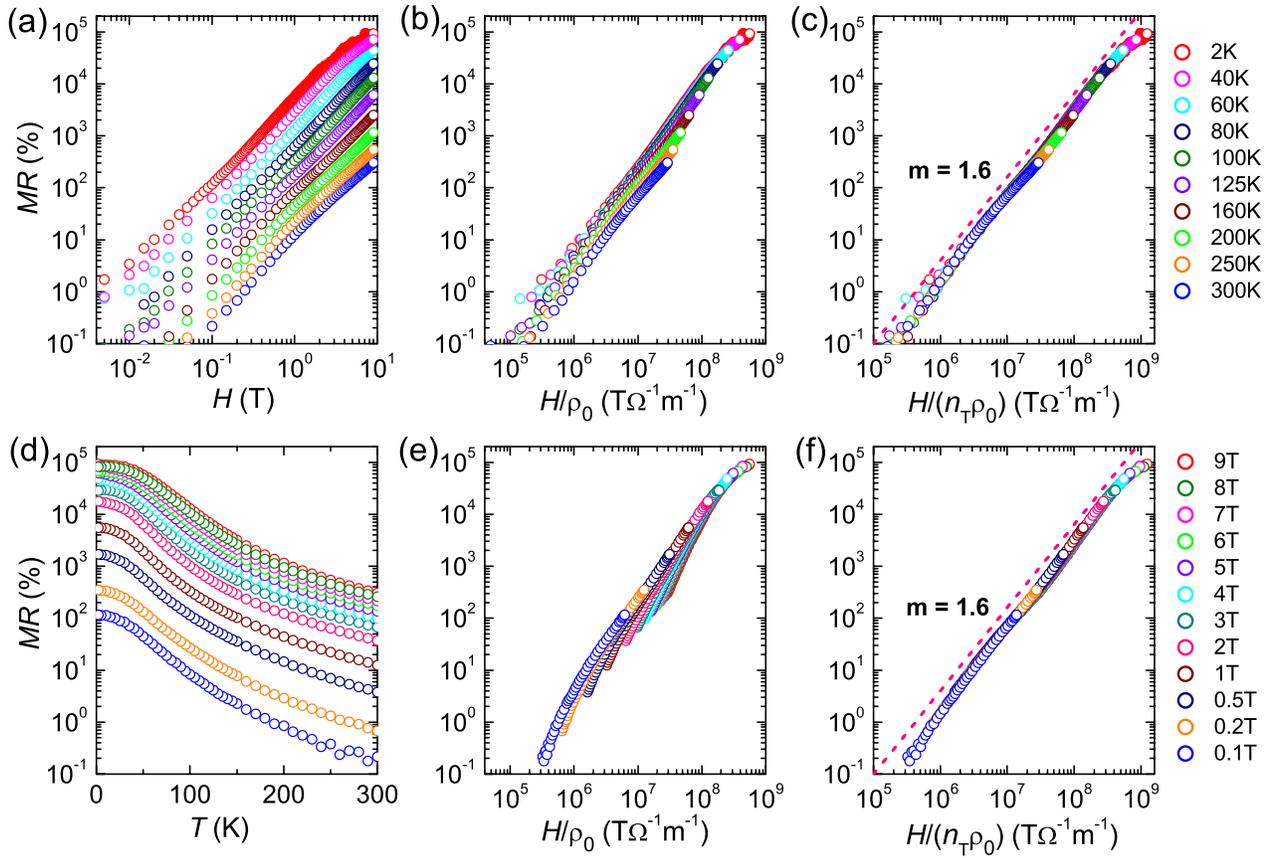

**Figure 3**

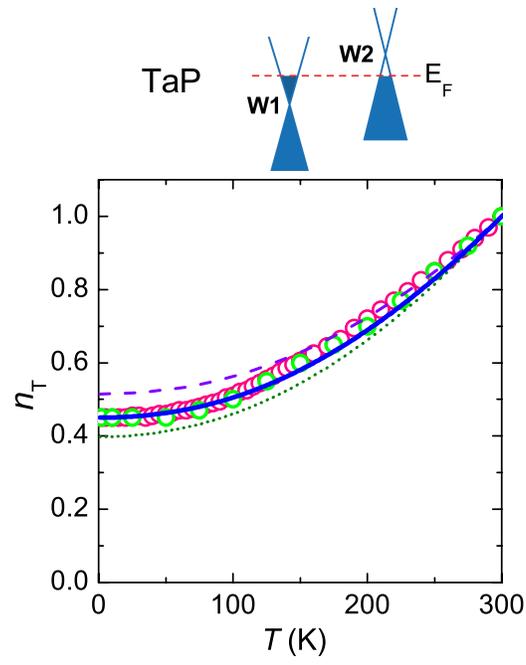

**Figure 4**

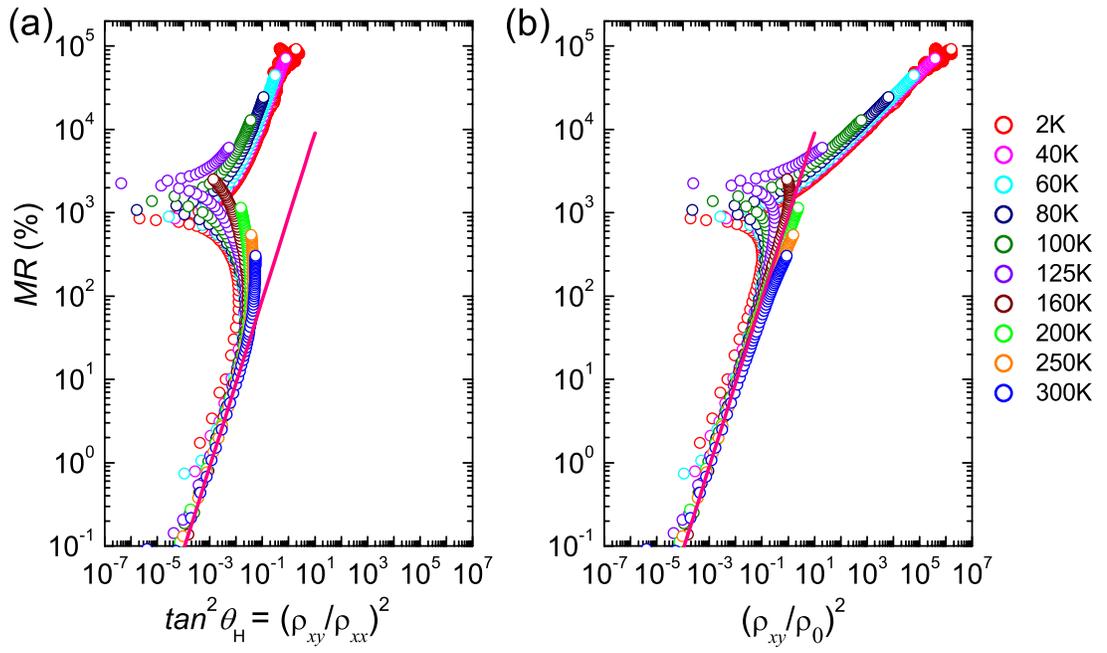

**Figure 5**

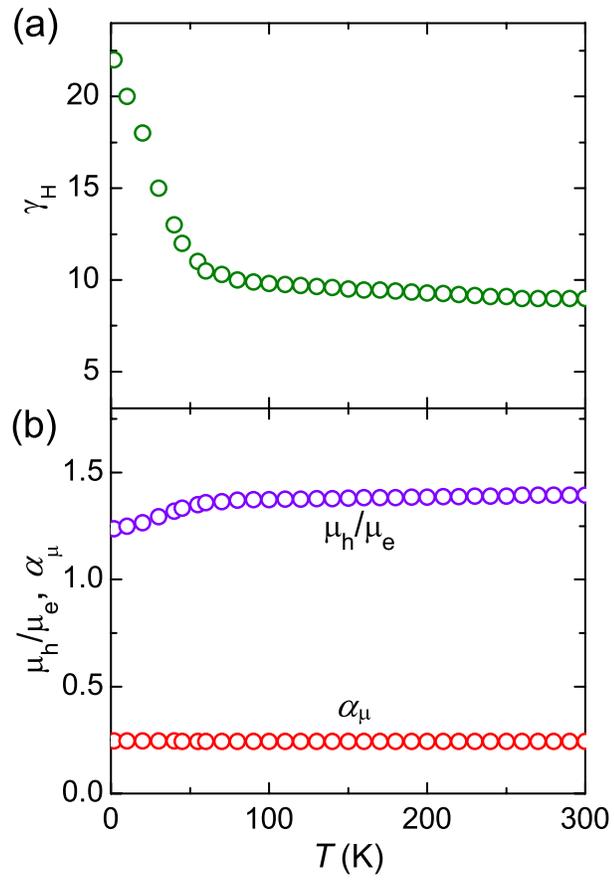

**Figure 6**

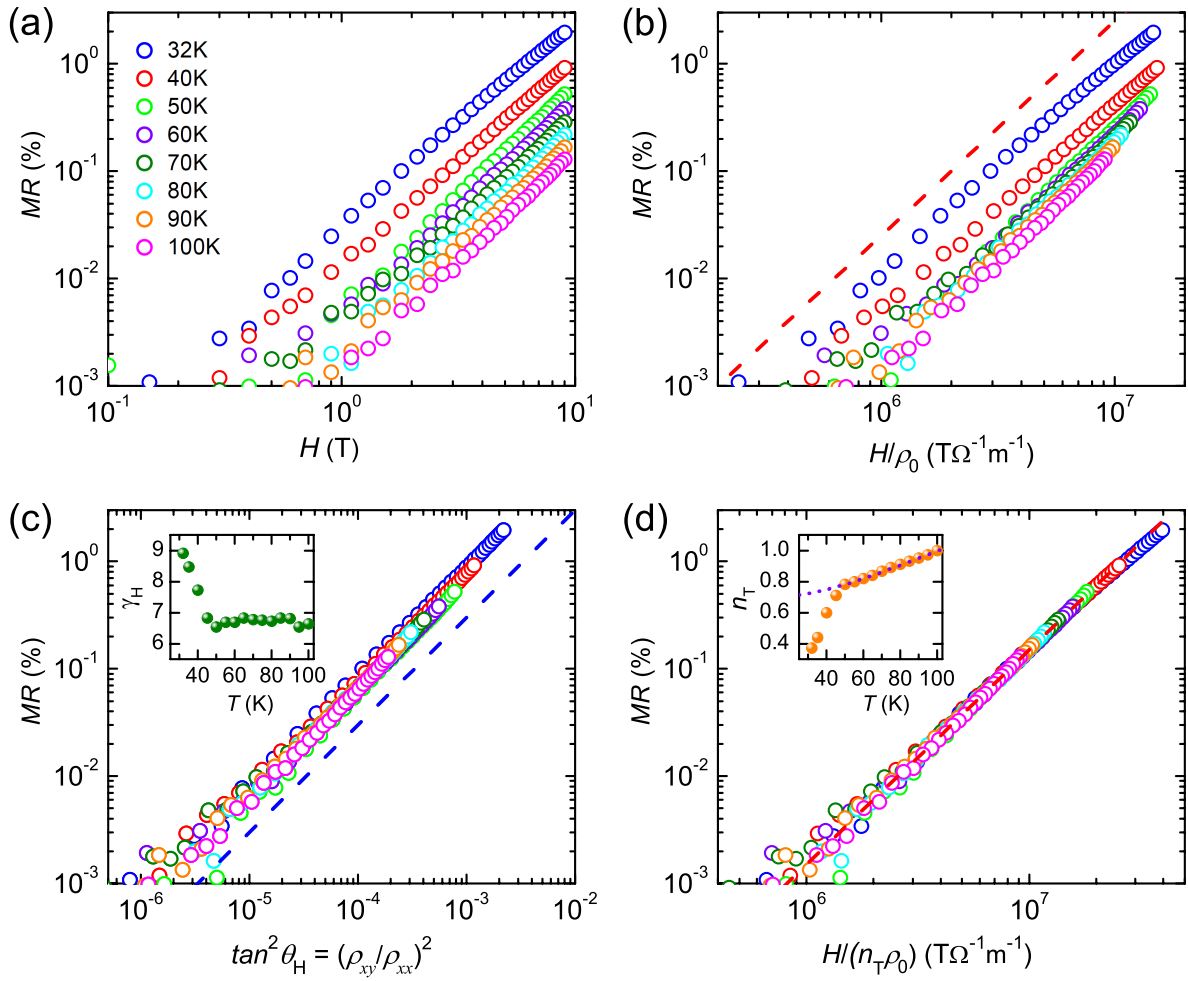

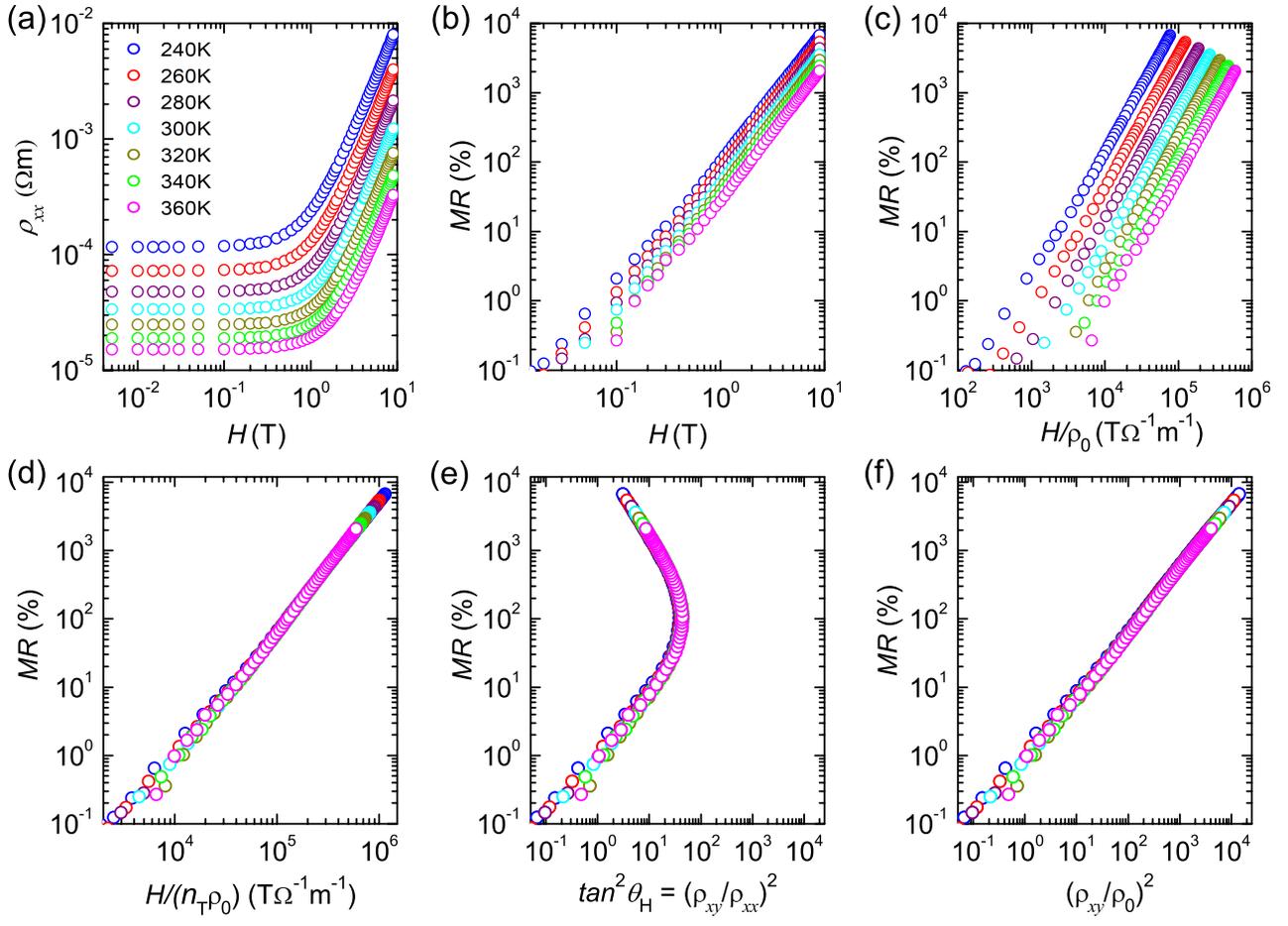

# Supplementary Information:

**Extended Kohler's Rule of Magnetoresistance**

Jing Xu et al.

**Equations for calculating the magnetoresistivities of a two-band system[1,9]**

$$\rho_{xx}(H) = \frac{1}{e} \frac{(n_h\mu_h + n_e\mu_e) + (n_h\mu_e + n_e\mu_h)\mu_h\mu_e H^2}{(n_h\mu_h + n_e\mu_e)^2 + (n_h - n_e)^2 \mu_h^2 \mu_e^2 H^2} \tag{S1}$$

$$\rho_{xy}(H) = \frac{H}{e} \frac{(n_h\mu_h^2 - n_e\mu_e^2) + (n_h - n_e)\mu_h^2 \mu_e^2 H^2}{(n_h\mu_h + n_e\mu_e)^2 + (n_h - n_e)^2 \mu_h^2 \mu_e^2 H^2} \tag{S2}$$

**Scaling behavior of magnetoresistance in an anisotropic single-band system with low carrier mobility**

We start with the carrier's equation of motion $\dot{\boldsymbol{v}} = -e(\boldsymbol{E} + \boldsymbol{v} \times \boldsymbol{H})/m^* - \boldsymbol{v}/\tau$, where $e, \boldsymbol{v}, m^*$ and $\tau$ are the charge, velocity, effective mass, and relaxation time of the carrier and $\boldsymbol{E}$ and $\boldsymbol{H}$ are the electric field and magnetic field, respectively[48]. For a single-band system with an anisotropic Fermi surface in $\boldsymbol{H} \parallel \boldsymbol{z}$, the effective mass is $m^* = \begin{pmatrix} m^*_{xx} & 0 \\ 0 & m^*_{yy} \end{pmatrix}$, leading to

$$-\frac{e}{m^*_{xx}}(E_x - v_y H) - \frac{v_x}{\tau} = 0 \qquad -\frac{e}{m^*_{yy}}(E_y + v_x H) - \frac{v_y}{\tau} = 0$$

They can be re-written as:

$$\mu_x(E_x - v_y H) + v_x = 0 \qquad \mu_y(E_y + v_x H) + v_y = 0$$

with $\mu_x = e\tau/m^*_{xx}$ and $\mu_y = e\tau/m^*_{yy}$

Thus, we obtain

$$v_x = (\mu_x E_x - \mu_x \mu_y H E_y)/(1 + \mu_x \mu_y H^2)$$

$$v_y = (\mu_y E_y + \mu_x \mu_y H E_x)/(1 + \mu_x \mu_y H^2)$$

Together with the definitions of $j_x = nev_x$ and $j_y = nev_y$ as well as $j_x = (\sigma_{xx} E_x + \sigma_{xy} E_y)$ and $j_y = (\sigma_{yx} E_x + \sigma_{yy} E_y)$ we can express the conductivities as

$$\sigma_{xx} = ne\mu_x/(1+\mu_x\mu_y H^2)$$

$$\sigma_{yy} = ne\mu_y/(1+\mu_x\mu_y H^2)$$

$$\sigma_{xy} = ne\mu_x\mu_y H/(1+\mu_x\mu_y H^2)$$

For clarity in discussions below we re-write them as

$$\sigma_{xx} = ne\mu_x/(1+\kappa\mu_x^2 H^2) \tag{S3}$$

$$\sigma_{yy} = ne\kappa\mu_x/(1+\kappa\mu_x^2 H^2) \tag{S4}$$

$$\sigma_{xy} = ne\mu_x^2 H/(1+\kappa\mu_x^2 H^2) \tag{S5}$$

where $\kappa = \mu_y/\mu_x$.

Considering the conversions of magnetoresistivities from magnetoconductivities, $\rho_{xx} = \sigma_{yy}/(\sigma_{xx}\sigma_{yy} + \sigma_{xy}^2)$ and $\rho_{xy} = \sigma_{xy}/(\sigma_{xx}\sigma_{yy} + \sigma_{xy}^2)$, we obtain $\rho_{xx} = 1/(ne\mu_x)$ and $\rho_{xy} = H/(ne)$. That is, a single band material would have no magnetoresistance if all carriers moving in the same direction have the same mobility.

However, carriers at energy levels near E_F also contribute to the conductance[1]. Thus, the carriers have a distribution of $n(\mu_x)$. Assuming $g = dn/d\mu_x$, we can have:

$$\sigma_{xx} = e\int_{\mu_L}^{\mu_H} \frac{g\mu_x d\mu_x}{1+\kappa\mu^2 H^2} \qquad \sigma_{yy} = e\int_{\mu_L}^{\mu_H} \frac{g\kappa\mu_x d\mu_x}{1+\kappa\mu_x^2 H^2} \qquad \sigma_{yx} = e\int_{\mu_L}^{\mu_H} \frac{g\kappa\mu_x^2 H d\mu_x}{1+\kappa\mu_x^2 H^2}$$

where $\mu_H$ and $\mu_L$ are the upper and lower limits of $\mu_x$, respectively. $g$, $\mu_H$ and $\mu_L$ satisfy the relationship $n = \int_{\mu_L}^{\mu_H} g d\mu_x$.

If the mobility is low such that $\mu_x^2 H^2 \ll 1$, we obtain $\sigma_{xx} \approx (\sigma_0 - aH^2)$ and $\sigma_{yx} \approx bH$

with $\quad \sigma_0 = e\int_{\mu_L}^{\mu_H} g\mu_x d\mu_x \qquad a = e\int_{\mu_L}^{\mu_H} g\kappa\mu_x^3 d\mu_x \qquad b = e\int_{\mu_L}^{\mu_H} g\kappa\mu_x^2 d\mu_x$

Since $\sigma_{xy} \ll \sigma_{xx}$ at low mobilities, the magnetoresistivities can be expressed as:

$$\rho_{xx} \approx \rho_0(1 + a\rho_0 H^2) \tag{S6}$$

$$\rho_{xy} \approx \rho_0^2 bH \tag{S7}$$

where $\rho_0 = 1/\sigma_0$ is the zero-field resistivity. From Eq.S6 we can obtain a parabolic magnetoresistance, i.e., $MR \approx a\rho_0 H^2$, where $MR = (\rho_{xx} - \rho_0)/\rho_0$. Combing Eq.S6 and Eq.S7 we have

$$MR \approx \gamma_H^*(\rho_{xy}/\rho_0)^2 \tag{S8}$$

with $\gamma_H^* = a/(\rho_0 b^2)$. Since $\rho_{xx} \approx \rho_0$ at low carrier mobility, Eq.S8 can be expressed as:

$$MR \approx \gamma_H^* tan^2\theta_H \qquad (S9)$$

Eq.S8 and Eq.S9 indicate that a low-mobility anisotropic single-band system can have similar scaling forms as those (Eq.5 and Eq.6) of a two-band system discussed in the text.

We can also re-write Eq.S6 as:

$$MR \approx \alpha_\mu^* [H/(n\rho_0)]^2 \qquad (S10)$$

where $\alpha_\mu^* = an^2\rho_0^3$. That is, the magnetoresistance of an anisotropic single-band system will follow the extended Kohler's rule Eq.1 if $\alpha_\mu^*$ is temperature independent. It may seem to be contradictory to claim that $\alpha_\mu^* (= an^2\rho_0^3)$ has no temperature dependence while $n$ can be temperature dependent. As demonstrated below, all $n$ in $\alpha_\mu^*$ get cancelled out because $a \sim n$ and $\rho_0 \sim 1/n$.

As an example, we consider a case that $n$ is equally distributed among $\mu_H \leq \mu_x \leq \mu_L$. In this case, $g = n/(\mu_H - \mu_L)$; $\rho_0 = 2/[en(\mu_H + \mu_L)]$; $a = en\kappa(\mu_H^4 - \mu_L^4)/[4(\mu_H - \mu_L)]$; resulting in

$$\alpha_\mu^* = 2\kappa/e^2 [1 + (\mu_L/\mu_H)^2]/[1 + \mu_L/\mu_H]^2 \qquad (S11)$$

That is, $\alpha_\mu^*$ is indeed temperature independent because $\mu_H$ and $\mu_L$ have the same temperature behavior (if the temperature does not induce a change in the anisotropy).

Likewise, we obtain a temperature independent $\gamma_H^*$:

$$\gamma_H^* = (9/8\kappa)[1 + (\mu_L/\mu_H)^2][1 - (\mu_L/\mu_H)^2]^2/[1 - (\mu_L/\mu_H)^3]^2 \qquad (S12)$$

**Scaling behavior of magnetoresistance in systems with two or more bands**

In a material with more than one bands, the magnetoconductivities for each band are $\sigma_{xx}^i = en_i\mu_i/[1 + (\mu_i H)^2]$ and $\sigma_{yx}^i = en_i\mu_i^2 H/[1 + (\mu_i H)^2]$, where $n_i$ and $\mu_i$ are the carrier density and mobility of the i[th] band. The total magnetoconductivities are $\sigma_{xx} = \sum_i \sigma_{xx}^i$ and $\sigma_{yx} = \sum_i \sigma_{yx}^i$. The magnetoresistivities can be calculated as $\rho_{xx} = \sigma_{xx}/(\sigma_{xx}^2 + \sigma_{yx}^2)$ and $\rho_{yx} = \sigma_{yx}/(\sigma_{xx}^2 + \sigma_{yx}^2)$.

Here, we derive the scaling behavior of the magnetoresistance MR at $\mu_i H \ll 1$, where $\sigma_{xx}^i \approx en_i\mu_i[1 - (\mu_i H)^2]$ and $\sigma_{yx}^i \approx en_i\mu_i^2 H$. In this case, we have

$$\rho_{xx} \approx 1/\sigma_{xx} = 1/\sum_i\{en_i\mu_i[1 - (\mu_i H)^2]\} \qquad (S13)$$

$$\rho_{yx} \approx \sigma_{yx}/\sigma_{xx}^2 = \sum_i(en_i\mu_i^2 H)/[\sum_i(en_i\mu_i)]^2 \qquad (S14)$$

From Eq.S13 we have $MR \approx [\sum_i(n_i\mu_i^3)/\sum_i(n_i\mu_i)]H^2$. Thus, we obtain,

$$MR = [H/(n_T\rho_0)]^2 \qquad (S15)$$

with

$$n_T = e[\sum_i(n_i\mu_i)]^{3/2}/[\sum_i(n_i\mu_i^3)]^{1/2} \qquad (S16)$$

Meanwhile we also have

$$MR \approx \gamma_H^{**}(\rho_{xy}/\rho_0)^2 \qquad (S17)$$

with $\gamma_H^{**} = \sum_i(n_i\mu_i^3)\sum_i(n_i\mu_i)/[\sum_i(n_i\mu_i^2)]^2$ and $\rho_0 = 1/\sum_i(en_i\mu_i)$.

The scalings of Eq.S15 and Eq.16 can be valid at high fields in materials with low mobilities or at low fields in those with high mobilities, as long as $\mu_i H \ll 1$.

Eq.S16 indicates that $n_T$ represents temperature dependence of the carrier density, i.e., $n_T \sim f_n(T)$, if the densities/mobilities from different bands have the same or similar temperature dependence, i.e., $n_i \approx n_i^0 f_n(T)$ and $\mu_i \approx \mu_i^0 f_\mu(T)$.

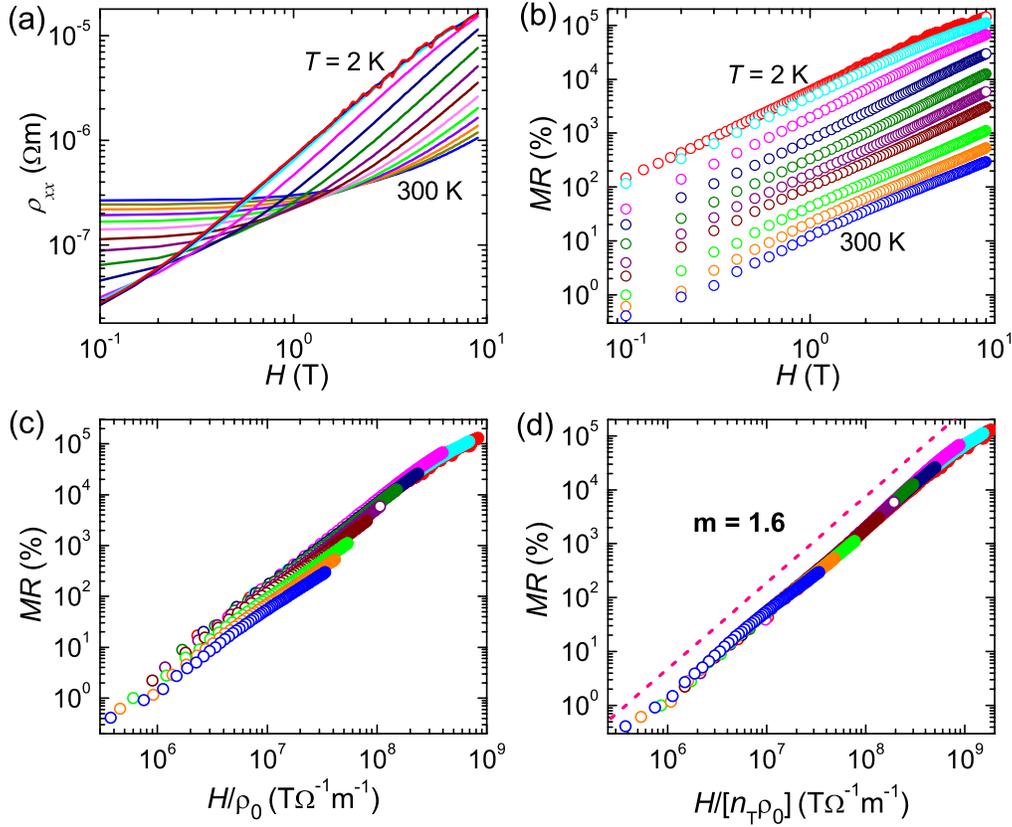

Fig.S1. Magnetoresistances and their scaling behavior in TaP (sample TP2). (a) Magnetic field dependence $\rho_{xx}(H)$ measured at temperatures of $T$ = 2 K, 10 K, and from 25 K to 300 K in intervals of 25 K. (b) Magnetic field of the MRs derived from data in (a) at $T$ = 2 K, and from 25 K to 150 K in intervals of 25 K, and at 200 K, 250 K and 300 K. (c) Kohler's rule scaling of the data in (b). (d) Extended Kohler's rule scaling of the data in (b). Same symbols are used in (b), (c) and (d). The dashed line in (d) indicates a power-law relationship with an exponent of m = 1.6.

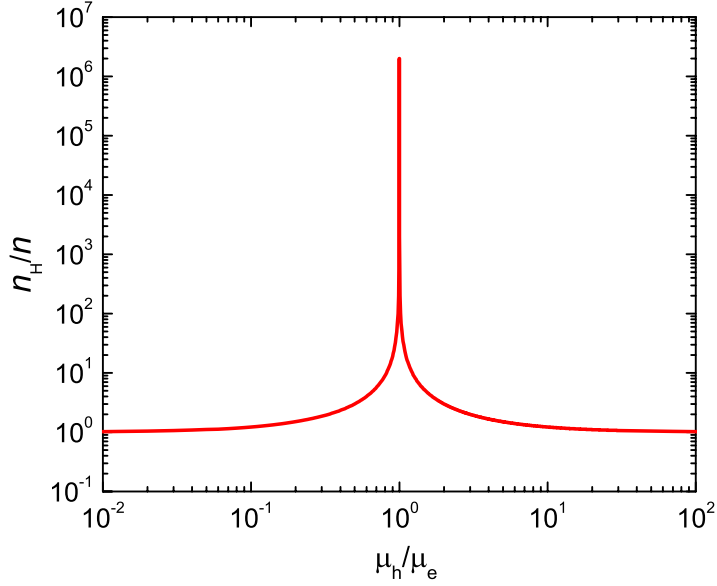

Fig.S2. Carrier density derived from the Hall effect measurements in a compensated two-band system. In transport experiments, the Hall effect can be employed to determine the densities of carriers in single band materials from the slopes of the linear Hall resistivity $\rho_{xy}(H)$ curves. When a two-band system is compensated, i.e., $n_e = n_h = n$, Eq.S2 also yields a linear field dependence of the Hall resistivity $\rho_{xy} = H/(en_H)$ with $n_H = n|(\mu_h/\mu_e + 1)/(\mu_h/\mu_e - 1)|$. In this case it is possible to obtain a carrier density $n_H$ from the Hall resistivity. However, the derived values $n_H$ can deviate significantly from the true electron and hole density $n$ in the material, depending on the ratio $\mu_h/\mu_e$ of the electron and hole mobilities $\mu_e$ and $\mu_h$ respectively. For example, 60K YBa$_2$Cu$_3$O$_{7-\delta}$, La$_2$Sr$_x$CuO$_4$, and TaP have the values of $\mu_h/\mu_e$= 2.2, $\mu_h/\mu_e$= 1.31 and $\mu_h/\mu_e$= 1.24-1.39, respectively (see discussions in the text), corresponding to $n_H/n$ = 2.8, $n_H/n$ = 7.45, and $n_H/n$ = 6.13-9.33 if they are compensated two-band materials.

These results indicate that $n_H$ determined using $\rho_{xy}(H)$ can differ from the true carrier density $n$ in the material even if $\rho_{xy}(H)$ shows a linear relationship.

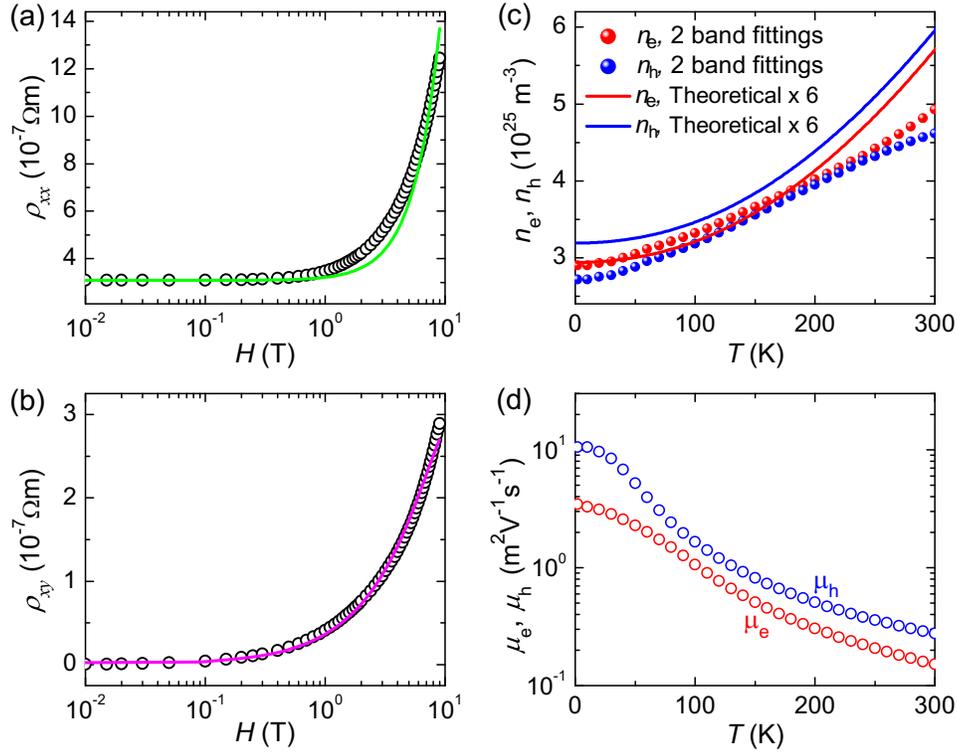

Fig.S3. Two-band model analysis of the magnetoresistivities in TaP (sample TP1). (a) and (b) Experimental $\rho_{xx}(H)$ and $\rho_{xy}(H)$ and the fits to Eq.S1 and Eq.S2, respectively. Open circles are data taken at $T$ = 300 K and lines are fits. (c) and (d) Derived densities and mobilities of the electrons and holes (symbols). The derived densities are nearly 6 times larger than the calculated values (lines) from Eq.2.

Currently, the only method to extract information on the carrier densities in a semimetal is to use the two-band model to fit either magnetoresistivities $\rho_{xx}(H)$ and $\rho_{xy}(H)$[5,10,13,26] or the converted magnetoconductivities $\sigma_{xx}(H)$ $[= \rho_{xx}/(\rho_{xx}^2 + \rho_{xy}^2)]$ and $\sigma_{xy}(H)$ $[= \rho_{xy}/(\rho_{xx}^2 + \rho_{xy}^2)]$ (Ref.14). Such an analysis with four fitting parameters (the densities and mobilities of electrons and holes $n_e$, $n_h$, $\mu_e$ and $\mu_h$) can result in unexpected outcomes. For example, in materials where Kohler's rule holds, i.e., $n_T$ in Eq.1 should be close to 1 and insensitive to temperature, the derived carrier densities can double or triple when the temperature is increased from 2 K to 300 K[14,26]. Here we use the measured zero-field resistivity $\rho_0$ to reduce the number of fitting parameters to three, and simultaneously fit both $\rho_{xx}(H)$ and $\rho_{xy}(H)$. The fitted curves, particularly those to the $\rho_{xx}(H)$ data, deviate significantly from the experimental data. One possible reason is that the two-band model is valid only for an isotropic system and is not suitable to describe the magnetoresistivities of TaP that has anisotropic Fermi pockets. Fittings with the two-band model yields values of $n_e$ and $n_h$ that are close to each other and increase with temperature. However, they are nearly a factor of 6 larger than the theoretical values. Furthermore, the fittings generate larger $n_e$ than $n_h$, in contrast to those obtained from bandstructure calculations and Eq.2.

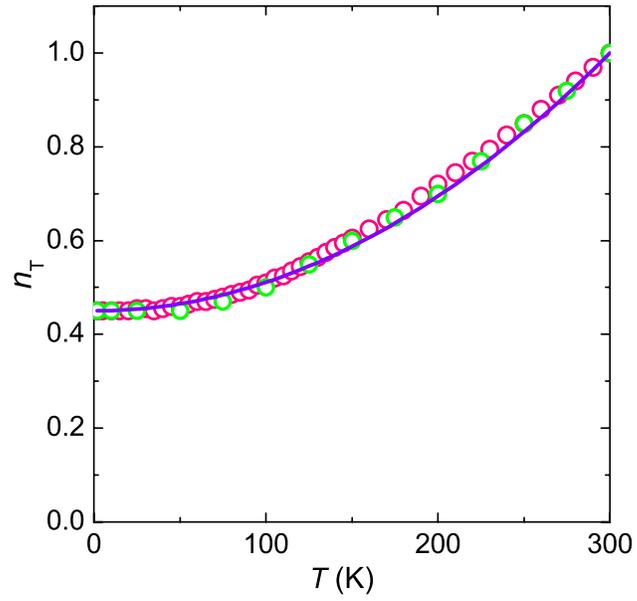

Fig.S4. Temperature dependence of $n_T$ obtained from extended Kohler's rule Eq.1 for TaP (sample TP1). Red and green circles are experimental results from sample TP1 and sample TP2, respectively. The purple curve is a power-law temperature dependence of $n_T \sim T^\alpha$ with $\alpha = 2$.

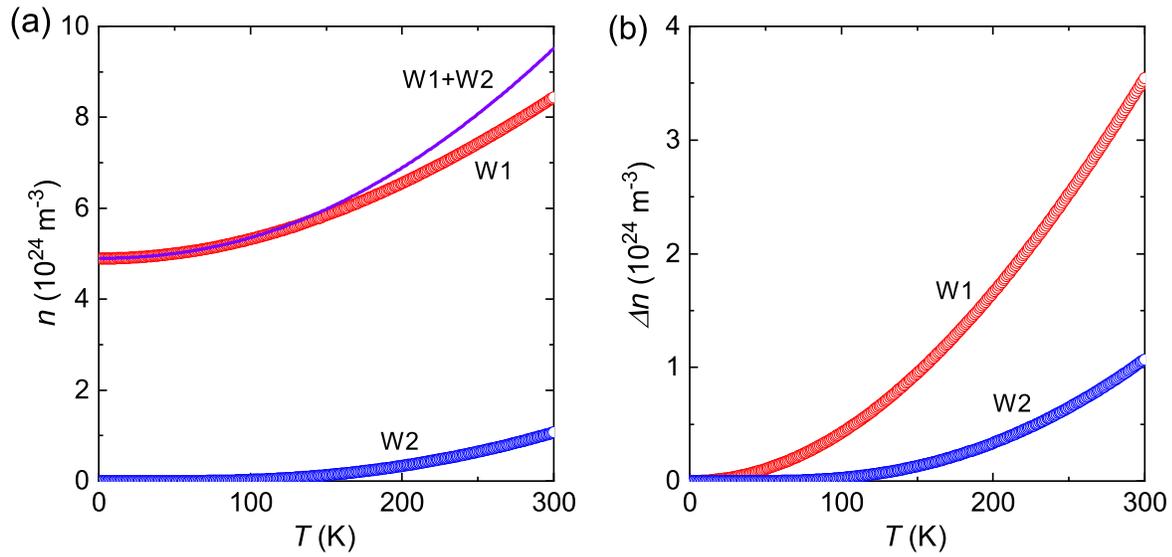

Fig.S5. Calculated temperature dependence of the electron density in TaP. (a) Electron densities calculated using Eq.2. We used theoretical values of -53.1 meV and 19.6 meV (relative to the Fermi level) for the energies of the Weyl nodes W1 and W2 and $n_e$ = 4.898×10$^{24}$ m$^{-3}$ and $n_h$ = 5.317×10$^{24}$ m$^{-3}$ for the electron and hole densities at $T$ = 0 K, respectively[44]. Red and blue circles are for electron densities of Weyl nodes W1 and W2, respectively and the purple line represents their sum. (b) Comparison of the thermally induced change of the electron density for W1 and W2, indicating that Weyl nodes W1 dominate the change over the entire temperature range.

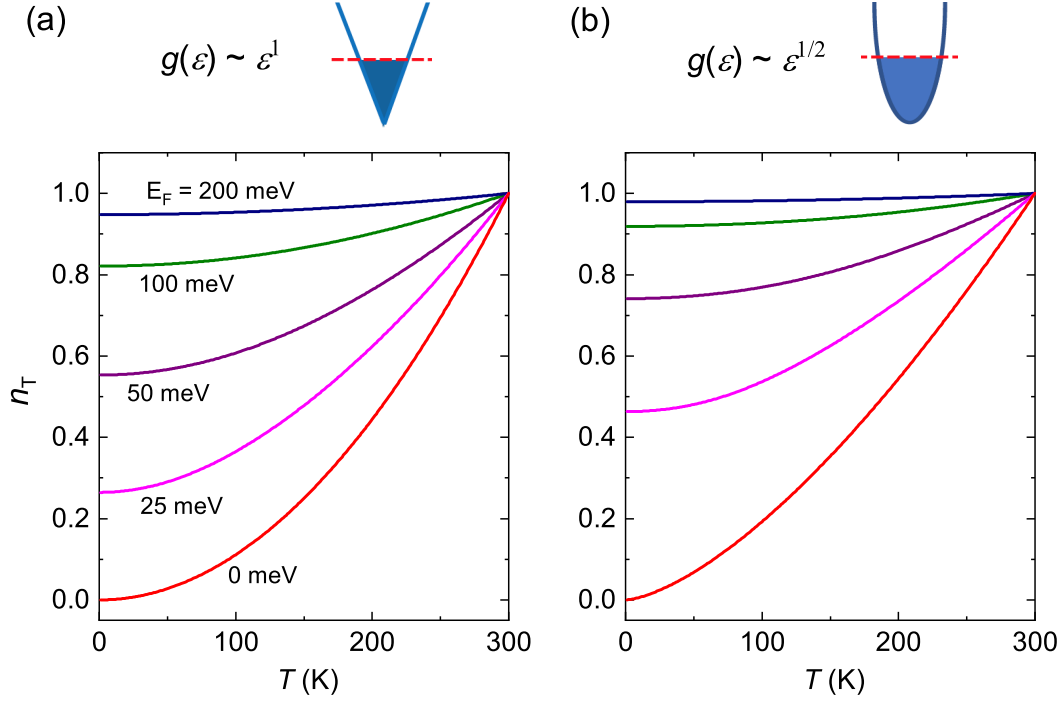

Fig.S6. Fermi level tuned temperature dependence of the electron density. (a) For Weyl semimetal TaP whose density of states (DoS) can be described as $g(\varepsilon) \sim \varepsilon^1$. (b) For metals with a DoS of $g(\varepsilon) \sim \varepsilon^{1/2}$. Schematics of the corresponding band structures are shown at the top of each panel, with the Fermi level indicated with a dashed red line. The legends in (b) are the same as those in (a). The values of $n_T$ are derived from Eq.2 and normalized to that at $T = 300$ K. The Fermi energies are relative values from the bottom of the conduction band. For simplicity and also as an estimate we used $g(\varepsilon) \sim \varepsilon^{\alpha}$ ($\alpha = 1/2$ and 1) at $\varepsilon$ from 0 to $\infty$ in the calculations, as typically done for free electrons. As discussed in the text for TaP, DoS can deviate from $g(\varepsilon) \sim \varepsilon^{\alpha}$ at high energies (e.g. $\varepsilon > 300$ meV). However, the contribution to the total electron density from the high energy levels (e.g. $\varepsilon > 300$ meV in TaP) is negligible (<0.1%).

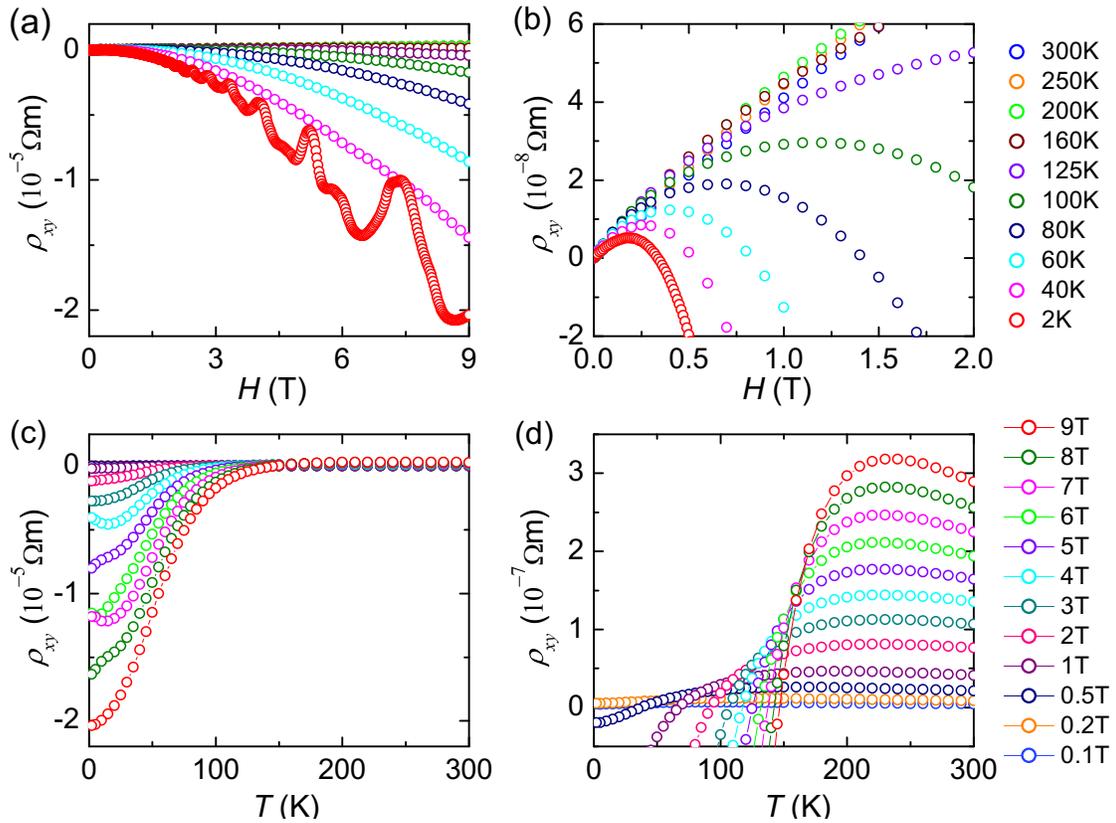

Fig.S7. Hall resistivities of TaP (sample TP1). (a) Magnetic field dependence of the Hall resistivity $\rho_{xy}(H)$ at various temperatures. (b) Expanded view of the data in (a). (c) Temperature dependence of the Hall resistivity $\rho_{xy}(T)$ at various magnetic fields. (d) Expanded view of the data in (c).

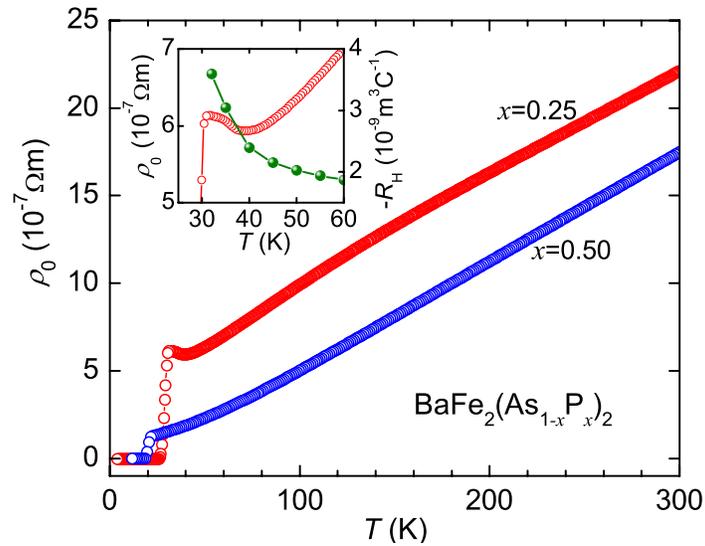

Fig.S8. Zero-field temperature dependence of the resistivity of BaFe$_2$(As$_{1-x}$P$_x$)$_2$ crystals with $x$ = 0.25 (sample PL) and 0.5 (sample PH). The inset presents an expanded view of the $\rho_0(T)$ curve for $x$ = 0.25 near the possible structural transition discussed by S. Kasahara et al. (Fig.1b of Ref.35]. The temperature dependence of the Hall coefficient $R_H$ (solid olive circles) in this sample does not show a discontinuity reported in crystals with lower levels of phosphorous while it does exhibit a significant upturn at $T \approx 45$ K.

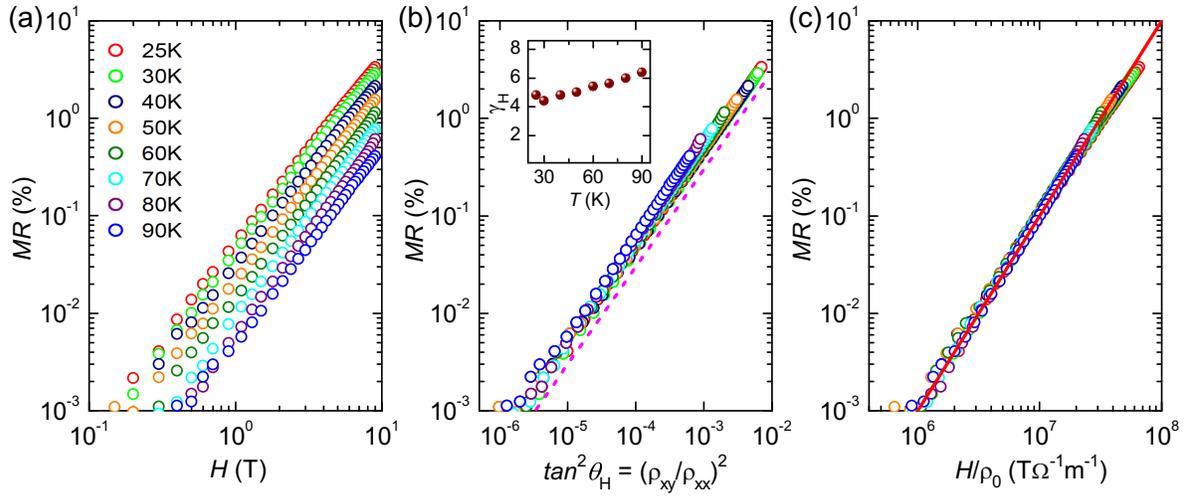

Fig.S9. Scaling behavior of the magnetoresistance of the BaFe$_2$(As$_{1-x}$P$_x$)$_2$ crystal with $x$ = 0.5 (sample PH). (a) $MR(H)$ curves at various temperatures. (b) Scaling according to Eq.3. The dashed straight magenta line that describes $MR = \gamma_H tan^2\theta_H$ with $\gamma_H = 3$ demonstrates the validity of Eq.3. $\gamma_H$ for each temperature is presented in the inset. (c) Scaling according to Kohler's rule. The straight red line represents $MR \sim (H/\rho_0)^2$. The same colored symbols are used in all panels.

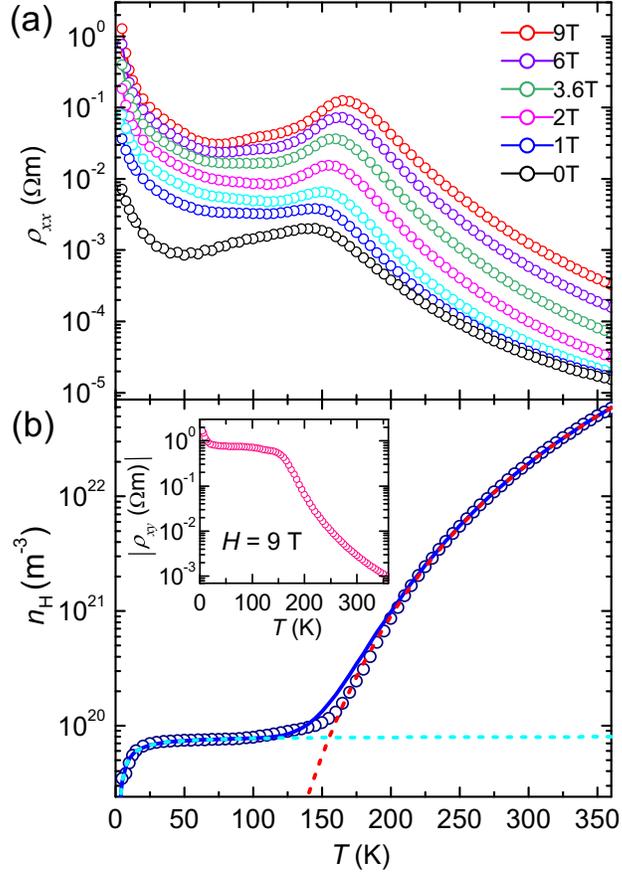

Fig.S10. Magnetoresistivity and Hall carrier density of InSb (sample IS). (a) Temperature dependence of the longitudinal resistivity $\rho_{xx}(T)$ at various magnetic fields. (b) Temperature dependence of the Hall carrier density $n_H = -1/[e\rho_{xy}]$ obtained from the Hall resistivity $\rho_{xy}(T)$ data in the inset.

The sample is a single crystal that is undoped but N-type due to residual impurities. The peaks in the $\rho_{xx}(T)$ curves in (a) and the bending in the $\rho_{xy}(T)$ curve in the inset of (b) at temperatures between 150 K and 175 K are due to a transition from the classical intrinsic state to the impurity dominated state, where quantum effects were reported[54]. The red dashed line in (b) is calculated with $n_H = 1.634 \times 10^{20} T^{3/2} e^{-E_g/2k_BT}$ using $E_g = E_{g0} + \alpha T^2/(\beta+T)$ with $E_{g0}$ = 235.2 meV, $\alpha$ = 0.35 meV /K and $\beta$ = 500 K. The cyan dashed line in (b) describes the impurity electron density $n^* = Ae^{-\Delta_0/k_BT}$ with $A = 8.1 \times 10^{19}$ m$^{-3}$ and $\Delta_0$ = 0.388 meV. The solid blue line is sum of $n_H$ and $n^*$. More discussions on the temperature dependence of the band gap are presented in the caption of Fig.S11b while those on the relationship between Hall density $n_H$ and intrinsic carrier density $n_i$ can be found in Fig.S12 and its caption.

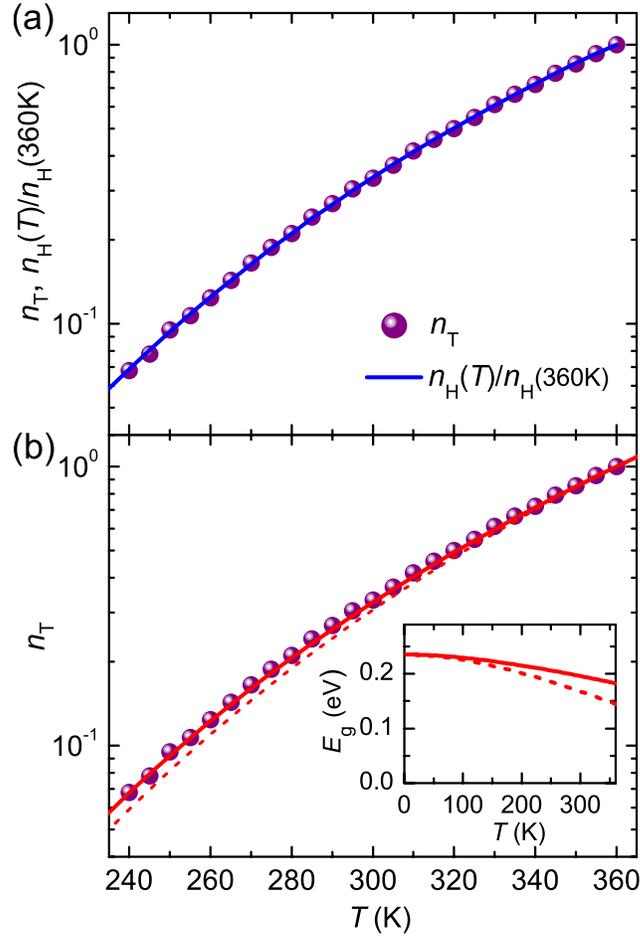

Fig.S11. Temperature dependence of $n_T$ in InSb (sample IS). (a) Comparison of the temperature dependence of $n_T$ with that of the Hall carrier density $n_H$ (Fig.S10b), which is normalized to the value at $T$ = 360 K. (b) Quantitative analysis of the $n_T$. Symbols are the same experimental data as those in (a). The red lines describe the expected temperature dependence of the carrier density thermally excited over a band gap $E_g$, i.e., $n_T \sim T^{3/2} e^{-E_g/2k_B T}$. The dashed line is derived using $E_g = E_{g0} + \alpha T^2/(\beta+T)$ with $E_{g0}$ = 235.2 meV, $\alpha$ = 0.6 meV /K and $\beta$ = 500 K from literature[53]. The solid line is a better fit using $\alpha$ = 0.35 while keeping $E_{g0}$ and $\beta$ the same. The inset of (b) shows a comparison of the corresponding band gaps.

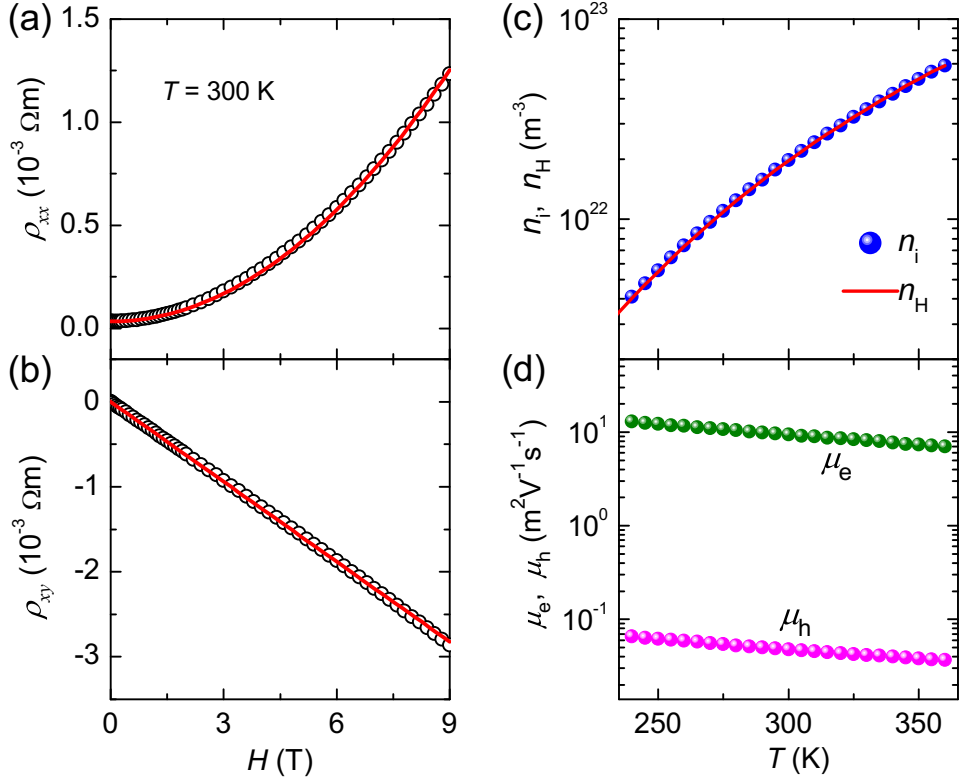

Fig.S12. Two-band model analysis of the magnetoresistivities in InSb (sample IS). (a) and (b) Fittings of both $\rho_{xx}(H)$ and $\rho_{xy}(H)$ curves at $T$ = 300 K simultaneously using Eq.S1 and Eq.S2. We neglected the contributions of the impurities at $T \geq 240$ K and used $n_e = n_h = n_i$. We also used the measured $\rho_0$ value and the relationship of $\rho_0 = e(n_e\mu_e+n_h\mu_h) = n_i e(\mu_e+\mu_h)$ to reduce the number of fitting parameters. This led us to only two free parameters of $n_i$ and $\mu_e$ while $\mu_h$ can be calculated from $\mu_h = \rho_0/n_i e - \mu_e$. (c) and (d) Temperature dependence of the derived parameters of $n_i$, $\mu_e$ and $\mu_h$. (c) indicates that the carrier density $n_i$ obtained from the two-band model analysis is indistinguishable from the Hall carrier density $n_H$ (Fig.S10b). This is consistent with the results in Fig.S2, which showed that the Hall carrier density, $n_H$, is equal to the carrier density, $n_i$, at very large ratios of $\mu_e/\mu_h$ (it changes slightly from 190 at 360 K to 199 at 240 K for the $\mu_e$ and $\mu_h$ in(d)).